\documentclass[pdflatex,sn-vancouver-num,iicol]{sn-jnl}

\usepackage{amsmath,amsfonts,amssymb}
\usepackage{textcomp}

\usepackage{algorithm}
\usepackage{algpseudocode}

\usepackage{graphicx}
\usepackage{pgfplots}
\pgfplotsset{compat=1.18} 
\usepackage{subcaption}   
\usepackage[table]{xcolor}
\usepackage{booktabs}

\usepackage{graphicx}
\usepackage{caption}
\usepackage{subcaption}
\usepackage{array}
\usepackage{tabularx}
\usepackage{booktabs}
\usepackage{multirow}
\usepackage[table]{xcolor}

\usepackage{dblfloatfix}

\usepackage{float}
\usepackage{url}
\usepackage{verbatim}
\usepackage{pifont}
\usepackage{placeins} 
\usepackage{balance}  

\usepackage{hyperref}
\hypersetup{
    colorlinks=false,
    pdfborder={0 0 0},
}

\newcommand{\green}[1]{\textcolor{green!60!black}{#1}}
\newcommand{\red}[1]{\textcolor{red}{#1}}
\definecolor{rowgray}{gray}{0.96}
\theoremstyle{thmstyleone}%
\theoremstyle{thmstyletwo}%

\theoremstyle{thmstylethree}%

\begin{document}

\title[Article Title]{When Popularity Fails: Fair and Equitable Long-Tail Recommendation with Fairness-Constrained Learning}


\author*[1]{\fnm{Subham} \sur{Raj}}\email{subham\_2221cs25@iitp.ac.in}

\author[1]{\fnm{Prateek}}\email{prateek\_2411ai40@iitp.ac.in}

\author[1]{\fnm{Sriparna} \sur{Saha}}\email{sriparna@iitp.ac.in}

\author[2]{\fnm{Brijraj} \sur{Singh}}\email{brijraj.singh@sony.com}

\author[2]{\fnm{Niranjan} \sur{Pedanekar}}\email{niranjan.pedanekar@sony.com}

\affil[1]{\orgname{Indian Institute of Technology Patna}, 
\orgaddress{\city{Patna}, \postcode{801103}, \state{Bihar}, \country{India}}}

\affil[2]{\orgname{Sony Research India}, 
\orgaddress{\city{Bangalore}, \postcode{560048}, \state{Karnataka}, \country{India}}}


\abstract{Recommendation systems play a critical role in modern digital platforms but often exhibit strong popularity bias, where a small subset of highly popular items dominates user exposure while the majority remain underrepresented. This imbalance limits content diversity and raises fairness concerns, as items from minority creators or niche communities frequently fall in the long tail. Existing long-tail and fairness-aware methods are typically developed independently and address bias only after representations have been learned from skewed interaction logs. We propose Fairness-Aware Biased User History Synthesis (FA-BUHS), a framework that integrates fairness constraints directly into the user history synthesis process, intervening before biased representations can form. Our fairness-constrained sampling strategy generates synthetic interaction histories that balance long-tail exposure with demographic group representation, enabling the model to learn user embeddings from a more equitable distribution. We further incorporate conditional adversarial debiasing to remove demographic leakage from latent embeddings while preserving collaborative filtering signals. Evaluated on MovieLens-1M and BookCrossing against long-tail, causal, and fairness-aware baselines, FA-BUHS improves tail-item NDCG@10 by up to 69.8\% and demographic parity by over 100\%, showing that upstream fairness-aware synthesis can jointly address popularity bias and fairness without sacrificing recommendation quality.}

\keywords
{Fairness-Aware Learning, Popularity Bias Mitigation, User History Synthesis, Recommendation System}



\maketitle

\section{Introduction}

Recommender systems play a central role in modern digital platforms by guiding users through vast catalogs of content and products. Most large-scale recommendation models rely heavily on historical interaction logs to learn user preferences. However, these interaction logs are inherently biased. In particular, a small subset of highly popular items tends to dominate user interactions, while the majority of items in the catalog---the so-called long tail---receive limited exposure. As a result, recommendation models trained on such data often reinforce popularity bias, repeatedly promoting already popular items while suppressing less frequently observed content \cite{klimashevskaia2024survey, abdollahpouri2019managing}.

Beyond reducing catalog diversity, popularity bias also raises critical fairness concerns. Items associated with minority creators, niche communities, or emerging providers frequently fall within the long tail of the catalog. When recommendation algorithms systematically prioritize head items, these groups receive disproportionately low exposure, leading to inequitable outcomes \cite{ekstrand2018all, jin2023survey}. Consequently, addressing popularity bias is not merely a matter of improving recommendation diversity, but a fundamental requirement for ensuring fair and balanced content visibility.

Prior work has explored several strategies to mitigate these issues. Long-tail recommendation methods aim to improve the coverage of underrepresented items by adjusting the training objective, re-weighting item popularity, or augmenting user interaction histories \cite{qin2021survey, zhang2021model}. A recent foundational example is Biased User History Synthesis (BUHS) \cite{balasubramanian2024biased}, which generates synthetic user histories that emphasize tail items to improve long-tail recommendation performance. In parallel, fairness-aware recommender systems attempt to mitigate demographic bias through adversarial learning, fairness regularization, or post-hoc re-ranking algorithms that enforce group-level exposure constraints \cite{zehlike2017fa, resheff2018privacy}.

However, these two research directions have largely evolved independently. Long-tail recommendation methods primarily focus on improving item coverage but do not explicitly account for demographic or provider fairness; they may even inadvertently amplify demographic stereotypes when synthesizing user histories. Conversely, fairness-aware recommenders typically operate downstream during representation learning or final ranking, attempting to scrub bias \textit{after} the model has already been trained on biased interaction histories. As a result, these approaches fail to address the underlying interaction bias that shapes the learned user representations at their source.

This observation motivates a different approach: fairness constraints should be applied earlier in the recommendation pipeline, during the construction of the user interaction histories themselves. If the training data used to learn user representations is intrinsically balanced with respect to item groups, the resulting embeddings are less likely to encode systematic exposure bias.

To this end, we propose \textbf{Fairness-Aware Biased User History Synthesis (FA-BUHS)}, a framework that integrates fairness constraints directly into the process of user history synthesis. Building upon the BUHS paradigm, our method introduces a fairness-constrained sampling strategy that balances long-tail exposure with demographic group representation. By synthesizing user histories that forcefully include items from underrepresented groups, the model learns user embeddings from a more balanced, counterfactual interaction distribution rather than the biased observational logs.

In addition to this fairness-constrained sampling, we incorporate conditional adversarial debiasing to prevent demographic attributes from leaking into the latent representations. Unlike standard adversarial fairness methods that indiscriminately remove sensitive information, our conditional adversarial formulation preserves the collaborative filtering signals that are genuinely predictive of user-item interactions. This dual design enables the model to mitigate demographic bias while maintaining rigorous recommendation accuracy.

We evaluate the proposed FA-BUHS framework on two benchmark datasets---MovieLens-1M and BookCrossing---and compare it against a comprehensive suite of long-tail recommendation, causal, and fairness-aware baselines. Our results demonstrate that FA-BUHS significantly improves exposure fairness and long-tail recommendation quality while maintaining competitive ranking accuracy.

The main contributions of this work are summarized as follows:
\begin{itemize}
    \item \textbf{Upstream Fairness Intervention:} Unlike existing fairness-aware recommenders that correct bias during or after representation learning, our framework addresses interaction bias earlier in the pipeline. We apply fairness constraints during user history construction, before biased embeddings can form.
    \item \textbf{Fairness-Constrained History Synthesis:} We introduce a fairness-aware extension of biased user history synthesis that integrates demographic group constraints into the generative sampling process, providing a bounded improvement in exposure for underrepresented item groups (Section 4.4.1).
    \item \textbf{Conditional Adversarial Debiasing:} We propose a conditional adversarial learning mechanism that reduces demographic leakage from latent representations while preserving the collaborative filtering signals necessary for high-fidelity personalization.
    \item \textbf{Comprehensive Empirical Evaluation:} Extensive experiments on multiple benchmark datasets demonstrate that FA-BUHS achieves a strong trade-off between recommendation accuracy, long-tail coverage, and exposure fairness compared to state-of-the-art baselines.
\end{itemize}

\section{Related Work}

\subsection{Long-Tail Recommendation}
The long-tail problem in recommender systems arises from the highly skewed distribution of item interactions, where a small number of popular items dominate user engagement while the majority of items receive limited exposure \cite{klimashevskaia2024survey}. Early approaches addressed this challenge through re-ranking strategies, diversity-promoting objectives, and popularity-based regularization designed to increase the visibility of tail items \cite{steck2011item, abdollahpouri2019managing, karboua2022mitigation}.

More recent work has explored representation learning techniques that improve long-tail recommendation through data augmentation or structural modeling. Neural collaborative filtering models and graph-based recommenders have demonstrated strong performance by capturing high-order relationships in user-item interaction graphs \cite{wang2019neural, he2020lightgcn, wei2023meta, pu2025graph}. In particular, Biased User History Synthesis (BUHS) \cite{balasubramanian2024biased} introduces a foundational approach that generates synthetic user histories emphasizing tail items, allowing recommendation models to learn more informative representations for sparse interaction settings. While such methods improve item coverage and recommendation diversity \cite{qin2021survey, zhang2021model}, they typically focus on popularity bias without explicitly considering fairness across demographic or provider groups. As a result, long-tail methods may inadvertently reinforce other forms of bias present in the interaction data.

\subsection{Fairness in Recommender Systems}
Fairness-aware recommender systems aim to ensure equitable treatment of different stakeholders, including users, item providers, and demographic groups \cite{abdollahpouri2020multistakeholder, jin2023survey}. A large body of work has adapted fairness definitions from machine learning---such as demographic parity and equal opportunity---to recommendation settings \cite{li2023fairness, wu2023fairness}.

Existing approaches can generally be categorized into three groups. Post-processing methods, such as FA*IR and related re-ranking algorithms, modify the final recommendation list to satisfy fairness constraints \cite{zehlike2017fa}. Regularization-based methods incorporate fairness penalties directly into the learning objective, encouraging the model to reduce bias during training \cite{kamishima2012fairness, khedr2023certifair}. Finally, adversarial learning approaches use gradient reversal layers to remove sensitive demographic information from learned representations \cite{resheff2018privacy}. Although these techniques can reduce demographic bias, they typically operate downstream in the recommendation pipeline after the model has already learned representations from inherently biased interaction logs. Consequently, these methods often struggle to fully eliminate the structural bias embedded in the underlying training data without degrading system utility.

\subsection{Causal and Adversarial Debiasing}
More recent work has explored causal approaches to mitigate bias in recommender systems \cite{zhang2021causal, kher2025learning}. Methods based on inverse propensity scoring (IPS) attempt to correct exposure bias by re-weighting observed interactions according to estimated propensities \cite{schnabel2016recommendations, zhou2024counterfactual}. Similarly, causal representation learning frameworks attempt to disentangle user preferences from confounding factors such as popularity or platform exposure mechanisms \cite{chen2023bias}.

Adversarial learning has also emerged as a powerful tool for fairness-aware representation learning \cite{deldjoo2021survey}. By training discriminators that predict sensitive attributes from latent embeddings, models can be forced to obfuscate demographic information in the representation space \cite{xu2021robust, zhang2024towards}. However, unconditional adversarial debiasing may inadvertently destroy useful collaborative filtering signals when demographic attributes are naturally correlated with genuine user preferences.

\subsection{Positioning of Our Work}
Our framework differs from existing approaches by introducing fairness constraints during user history synthesis, rather than relying solely on representation scrubbing or post-ranking adjustments. By integrating fairness-aware sampling into the BUHS framework \cite{balasubramanian2024biased}, FA-BUHS constructs synthetic interaction histories that balance long-tail exposure with demographic group representation. This upstream intervention allows the model to learn user embeddings from a more equitable interaction distribution, targeting the interaction bias at its source rather than after it has propagated into learned representations. Furthermore, we combine this upstream synthesis with conditional adversarial debiasing, which selectively removes demographic leakage while preserving useful collaborative filtering signals. This dual-stage design enables FA-BUHS to address both popularity bias and demographic fairness within a unified recommendation framework.

Beyond fairness in recommendation specifically, related lines of work have examined value-driven user modeling in multi-agent recommendation settings \cite{chen2025multi} and the vulnerability of collaborative recommender systems to adversarially injected users \cite{yuan2023manipulating}, both of which raise complementary concerns about the reliability and manipulability of the interaction data recommender systems are trained on.

\section{Background and Notation}

We establish the mathematical foundation of our framework for fairness-aware biased user history synthesis (BUHS) in recommendation systems. This section introduces the graph-based problem formulation, dual-tower architecture, synthesis mechanism, and adversarial fairness module.

\subsection{Problem Formulation}

\textbf{User-Item Bipartite Graph.}  
We model user--item interactions as a bipartite graph 
\( G = (V, E) \), where the node set is 
\( V = U \cup I \), with users 
\( U = \{u_1, \ldots, u_{|U|}\} \) and items 
\( I = \{i_1, \ldots, i_{|I|}\} \). 
Each edge \( (u,i) \in E \) denotes that user \( u \) interacted with item \( i \), possibly with explicit (rating) or implicit (binary) feedback.

The graph is stored as an adjacency matrix:
\begin{equation}
A = \begin{bmatrix}
0_{|U| \times |U|} & R \\
R^T & 0_{|I| \times |I|}
\end{bmatrix},
\end{equation}
where \( R \in \{0,1\}^{|U|\times|I|} \) is the user--item interaction matrix, with 
\( R_{u,i} = 1 \) if \( (u,i) \in E \), and 0 otherwise.

\textbf{Long-Tail Categorization.}  
Items are partitioned into head \( I_h \) and tail \( I_t \), such that 
\( I_h \cap I_t = \varnothing \) and \( I_h \cup I_t = I \). 
An item \( i \) is in the tail set if its degree 
\( d_i = |\{u : (u,i)\in E\}| \leq \theta \), 
where \( \theta \) is a dataset-specific threshold, typically chosen so that tail items represent around 80\% of the catalog but only 20\% of interactions.

\subsection{Two-Tower Neural Architecture}

Our framework builds on the Two-Tower Neural Network (TTNN), widely adopted in large-scale recommender systems. The TTNN consists of two encoders:

\textbf{User tower:}
\begin{equation}
u = \phi_u(x_u; \Theta_u), \quad \phi_u : \mathcal{X}_u \rightarrow \mathbb{R}^d
\end{equation}
where \( x_u \) includes user ID, demographics, and contextual features, and \( \Theta_u \) are learnable parameters.

\textbf{Item tower:}
\begin{equation}
v = \phi_i(x_i; \Theta_i), \quad \phi_i : \mathcal{X}_i \rightarrow \mathbb{R}^d
\end{equation}
where \( x_i \) includes item ID, metadata, or multimodal content features.

The relevance score is computed as:
\begin{equation}
s_{u,i} = f(u,v),
\end{equation}
using either dot-product similarity (\( u^T v \)) or an MLP over concatenated embeddings.

\subsection{Tail-Biased User History Synthesis}

\textbf{Biased Sampling.}  
Given a user history $H_u = \{i : (u,i)\in E\}$, the original BUHS framework defines a tail-biased sampling distribution:
\begin{equation}
P(i|u) = \frac{\exp(-\log(d_i+1)/\tau)}{\sum_{j\in H_u}\exp(-\log(d_j+1)/\tau)}
\label{eq:base-sampling}
\end{equation}
where $d_i$ is the item degree and $\tau > 0$ controls the bias strength. A smaller $\tau$ increases the sampling preference for long-tail items, improving coverage.

\subsection{Background on Adversarial Debiasing}

To enforce fairness at the representation level, standard adversarial training adopts a gradient reversal layer (GRL):
\begin{equation}
\mathcal{R}_\lambda(x) = x \quad \text{(forward)}, \quad
\frac{\partial \mathcal{R}_\lambda}{\partial x} = -\lambda I_d \quad \text{(backward)},
\label{eq:grl_base}
\end{equation}
where $\lambda > 0$ controls the adversarial strength.

In standard unconditional adversarial debiasing, a discriminator $D$ attempts to predict a sensitive demographic attribute $a$ directly from the latent representation $z$. The model is optimized via a minimax game:
\begin{equation}
L_{\text{adv}} = \mathbb{E} \left[-\log D(a \mid \mathcal{R}_\lambda(z))\right]
\label{eq:unconditional_adv}
\end{equation}
While effective for general domain adaptation, applying this unconditional erasure in recommender systems risks destroying highly correlated, legitimate collaborative filtering signals alongside spurious biases.

\section{Proposed Methodology}

We propose a fairness-aware extension of Biased User History Synthesis (BUHS)\cite{balasubramanian2024biased} that addresses demographic bias and exposure inequity while preserving long-tail recommendation benefits. The original BUHS demonstrates through information-theoretic analysis that tail items provide higher pointwise mutual information for user characterization. Our FA-BUHS framework extends this principle by integrating fairness constraints during sampling, synthesizing user histories with fairness-aware models, applying adversarial debiasing, and providing comprehensive theoretical guarantees on fairness-utility trade-offs. 

\begin{figure*}
\centering
    \includegraphics[width=0.85\textwidth]{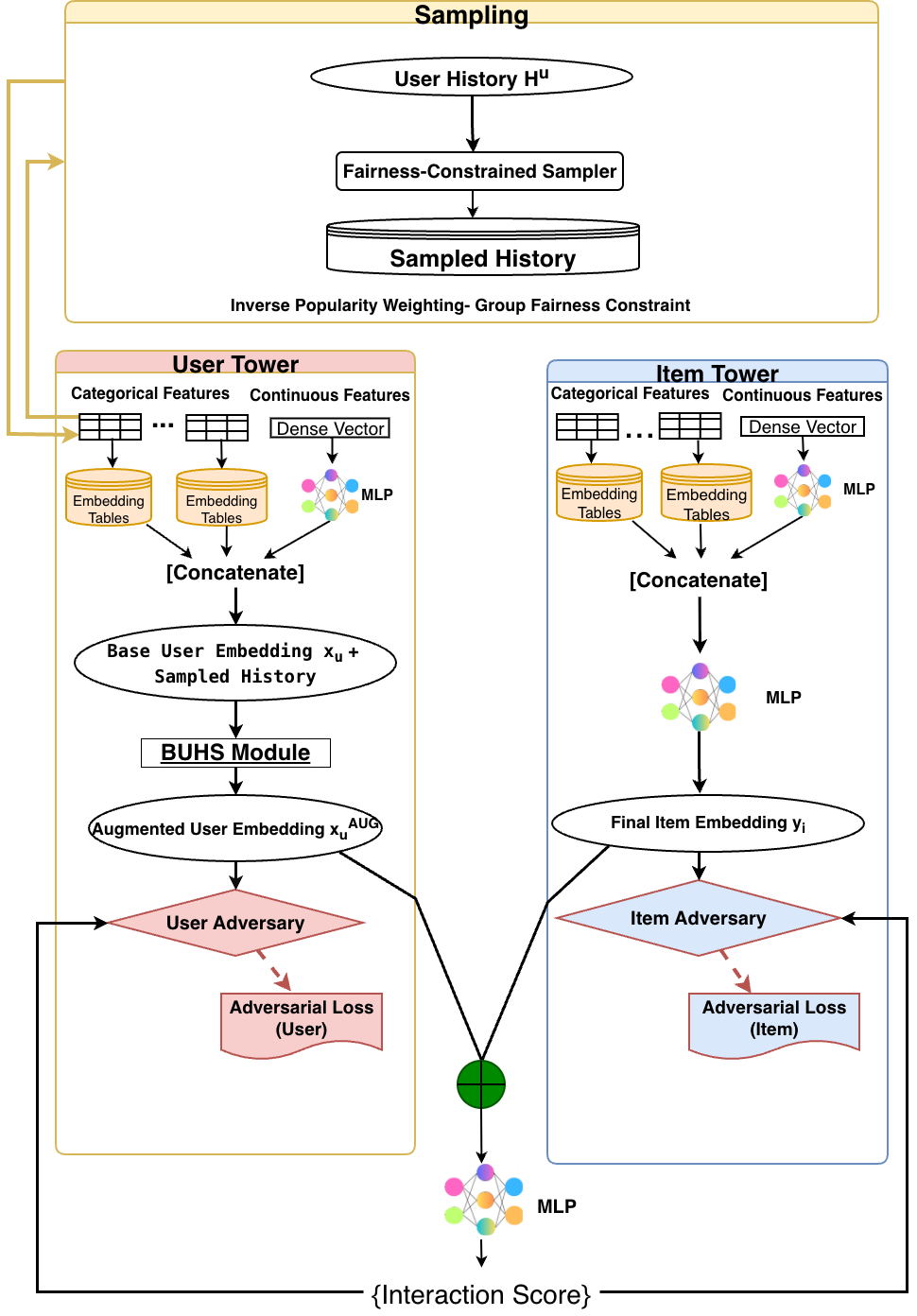}
    \caption{FA-BUHS architecture: (1) fairness-constrained sampling for balanced group representation, (2) dual-tower embedding learning, and (3) conditional adversarial debiasing to remove spurious demographic bias while preserving preference signals.}
    \label{fig:fabuhs_architecture}
\end{figure*}
\begin{figure}
    \centering
    \includegraphics[width=0.5\textwidth]{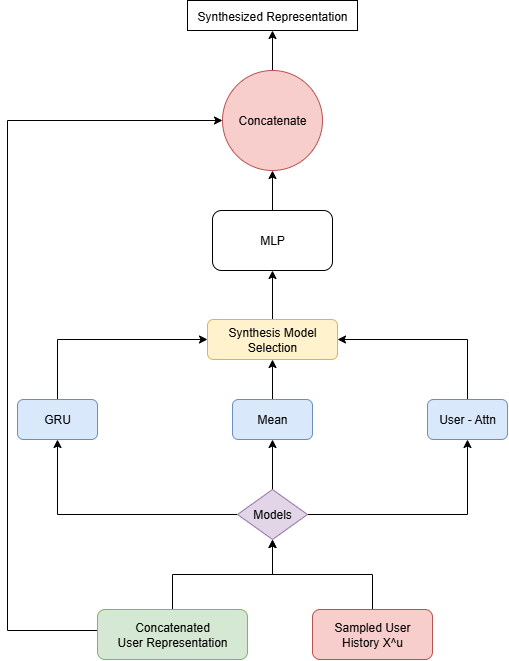}
    \caption{Architecture of the BUHS Synthesis Module in FA-BUHS. It supports three variants: (1) Mean Synthesis (element-wise mean of sampled items), (2) User-Attention Synthesis (scaled dot-product attention), and (3) GRU Synthesis (temporal sequence modeling). These enhance user representations while preserving BUHS benefits for long-tail recommendation.}
    \label{fig:buhs_detailed}
\end{figure}

The overall architecture is shown in Fig. \ref{fig:fabuhs_architecture} and the detailed architecture of BUHS Synthesis is shown in Fig. \ref{fig:buhs_detailed}.

\subsection{Fairness-Constrained Tail-Biased Sampling}
Building upon the original BUHS tail-biased sampling distribution (defined in Eq. \ref{eq:base-sampling}), we introduce group-level fairness constraints to ensure equitable representation across demographic groups while maintaining tail-focused benefits. 

Let $g(i)$ denote the function that maps an item $i$ to its respective demographic or genre group, and $\pi_{g(i)}$ denote its catalog prevalence. We define our fairness-aware sampling probability as:
\begin{equation}
P_{\mathrm{fair}}(i|u) = \frac{\exp\left(\frac{w_{g(i)} \cdot \log(d_i^{-1})}{\tau}\right)}{\sum_{j \in H_u} \exp\left(\frac{w_{g(j)} \cdot \log(d_j^{-1})}{\tau}\right)}
\label{eq:fair-sampling}
\end{equation}
where $d_i$ is the item degree, $\tau$ is a temperature parameter controlling the bias strength, and the group-specific fairness weight is:
\begin{equation}
w_{g(i)} = \frac{1-\alpha}{\pi_{g(i)} + \epsilon}
\label{eq:fair-weight}
\end{equation}
By embedding the fairness weight within the exponent of Eq.~\eqref{eq:fair-sampling}, 
it acts as a dynamic, group-specific temperature scalar. For underrepresented groups 
where $\pi_{g(i)}$ is small, $w_{g(i)} > 1$, which effectively lowers the temperature, 
thereby sharply amplifying the sampling probability of tail items from that minority 
group. Conversely, for overrepresented groups where $\pi_{g(i)}$ is large, $w_{g(i)} 
\leq 1$, which raises the effective temperature and suppresses their over-sampling. 
Because the weight $w_{g(i)}$ varies across items in the denominator summation 
according to their respective group affiliations $g(j)$, it cannot be factored out 
algebraically, preserving the integrity of the normalized distribution. The parameter 
$\alpha \in [0, 1)$ controls the fairness--accuracy trade-off, where larger values 
enforce stronger group-level equity, and $\epsilon > 0$ prevents division by zero for 
groups with negligible catalog prevalence.

\textbf{Lemma 1 (Tail Item Exposure Enhancement):}
Under fairness-constrained sampling with weights $w_{g(i)}$, the exposure increase for tail items in underrepresented groups is bounded by:
\begin{equation}
\frac{\mathrm{Exposure}_{\mathrm{fair}}(i,g)}{\mathrm{Exposure}_{\mathrm{original}}(i,g)} \geq \frac{1-\pi_g}{\pi_g + \epsilon}
\label{eq:lemma1}
\end{equation}
for items $i$ in underrepresented groups where $\pi_g < 0.5$. This lemma demonstrates that our weighting mechanism systematically increases visibility for tail items from minority groups, directly addressing the intersectional bias where minority demographics correlate with unpopular items.

The sampling is stratified across groups, ensuring both tail coverage and equitable representation across protected attributes such as gender, age, or geographic location.

\subsubsection{Causal Interpretation of Fairness-Constrained Sampling}
The FA-BUHS sampling mechanism can be formally interpreted through the lens of causal inference in recommendation systems. In standard pipelines, the observed user interaction history ($H_u$) is heavily confounded by platform exposure bias and popularity dynamics. Consequently, the baseline user representation ($x_u$) is estimated from a biased observational sample. 

By synthesizing an alternative, fairness-aware interaction history ($H_u^{\mathrm{fair}}$), FA-BUHS executes a structural causal intervention on the data generation process. From a Pearlian causal perspective, this procedure corresponds to the intervention $do(H_u = H_u^{\mathrm{fair}})$. This breaks the causal pathway through which historical popularity bias propagates into the user embeddings. 
Therefore, the resulting augmented user embedding ($x_u^{\mathrm{AUG}}$) is learned from an interventionally balanced counterfactual distribution rather than the biased observational data, framing FA-BUHS as an interpretable analogue to a causal intervention on the training distribution.

\subsection{Fairness-Aware User History Synthesis}
To construct fairness-preserving user embeddings that maintain the representational benefits of the original BUHS architecture, we introduce three complementary synthesis approaches, each enhanced with demographic bias correction mechanisms:

\paragraph{Mean Synthesis}
Aggregates sampled items through mean pooling with fairness regularization:
\begin{equation}
x_u^{\mathrm{AUG}} = W_{\mathrm{concat}} [x_u \| \mathrm{MEAN}(Y_{\mathrm{fair}})] + b
\end{equation}
where $Y_{\mathrm{fair}}$ represents the fairness-constrained sample set obtained through our stratified sampling strategy.

\paragraph{Attention-Based Synthesis}
Uses scaled dot-product attention with demographic correction to ensure that attention weights do not inadvertently encode protected attributes:
\begin{align}
A_{\mathrm{attn}} &= \mathrm{softmax}\left(\frac{Q K^T}{\sqrt{d_k}}\right) \\
h^{\mathrm{attn}} &= A_{\mathrm{attn}} V \\
x_u^{\mathrm{AUG}} &= W_{\mathrm{concat}} [x_u \| h^{\mathrm{attn}} \| f_{\mathrm{demo}}] + b
\end{align}
where $Q \in \mathbb{R}^{1 \times d_k}$ is the query matrix obtained by linearly projecting the base user embedding $x_u$. $K \in \mathbb{R}^{|Y_{\mathrm{fair}}| \times d_k}$ and $V \in \mathbb{R}^{|Y_{\mathrm{fair}}| \times d_k}$ are the key and value matrices, generated by applying distinct learned transformations to the embeddings of the sampled historical items $Y_{\mathrm{fair}}$. The scaling factor $\sqrt{d_k}$ prevents vanishing gradients. Finally, $f_{\mathrm{demo}}$ is a learned demographic context vector that counteracts spurious demographic correlations during the concatenation step.

\paragraph{Temporal GRU Synthesis}
Models ordered interaction histories while incorporating a fairness gate that regulates the influence of potentially biased temporal patterns:
\begin{align}
h_t, h_{\mathrm{final}} &= \mathrm{GRU}(Y_{\mathrm{sorted}}) \\
g_{\mathrm{fair}} &= \sigma(W_g h_{\mathrm{final}} + c_{\mathrm{demo}} + b_g) \\
x_u^{\mathrm{AUG}} &= W_{\mathrm{concat}} [x_u \| g_{\mathrm{fair}} \odot h_{\mathrm{final}}] + b
\end{align}
where $c_{\mathrm{demo}}$ encodes demographic context and $\odot$ denotes element-wise multiplication. The fairness gate $g_{\mathrm{fair}}$ learns to suppress biased temporal patterns while preserving genuine preference signals.

\textbf{Connecting Representation to Exposure Fairness:} It is critical to note how this synthesis translates into fair exposure in the final rankings. The synthesized representation ($x_u^{\mathrm{AUG}}$) integrates fairness-constrained samples ($Y_{\mathrm{fair}}$) with the base user embedding ($x_u$). Because the final ranking score is computed via the dot product ($s_{u,i} = {x_u^{\mathrm{AUG}}}^T v_i$), the contribution of the sampled tail items directly influences the similarity landscape of the user embedding. Specifically, if $Y_{\mathrm{fair}}$ contains items from underrepresented groups, the augmented embedding structurally shifts the user representation toward these items in the latent space. Consequently, the dot-product similarity for those specific minority items increases relative to their baseline values. Therefore, while the subsequent adversarial learning prevents demographic bias from re-emerging in the latent representations, it is this upstream fairness-aware sampling that actively reshapes the embedding geometry to redistribute exposure probabilities in the final ranking.

\subsection{Conditional Adversarial Debiasing}
To reduce the leakage of protected attributes while strictly maintaining task-relevant collaborative signals, we integrate conditional adversarial learning with gradient reversal layers. Instead of unconditionally erasing demographic data, our discriminators $D_u$ and $D_i$ are conditioned on the target interaction label $y_{u,i}$.

The total loss combines recommendation accuracy and conditional adversarial fairness objectives:
\begin{equation}
\begin{aligned}
L_{\mathrm{total}} 
&= L_{\mathrm{rec}} 
+ \lambda_u L_{\mathrm{cond\_adv}}(u, y_{u,i}) \\
&\quad + \lambda_i L_{\mathrm{cond\_adv}}(i, y_{u,i})
\end{aligned}
\end{equation}
where $\lambda_u, \lambda_i$ control the adversarial strength. The encoder is optimized to obfuscate demographics only when they do not contribute to predicting the true interaction $y_{u,i}$:
\begin{align}
L_{\mathrm{cond\_adv}}(u, y_{u,i}) &= \mathbb{E}_{u,i} \left[ -\log D_u (a_u | \mathcal{R}(x_u^{\mathrm{AUG}}), y_{u,i}) \right] \\
L_{\mathrm{cond\_adv}}(i, y_{u,i}) &= \mathbb{E}_{u,i} \left[ -\log D_i (a_i | \mathcal{R}(v_i), y_{u,i}) \right]
\end{align}
where $\mathcal{R}$ represents the gradient reversal operation. This guarantees that genuine collaborative filtering signals are insulated from the adversarial erasure process.

\subsection{Theoretical Insights and Bounding Analysis}
\textbf{Interpretation of Theoretical Analysis:} The following analysis provides theoretical intuition for the behavior of fairness-constrained sampling and adversarial representation learning. These derivations are not intended to serve as strict, global optimality guarantees. Instead, similar to many fairness-aware learning frameworks, these results should be interpreted as bounding analyses that characterize the expected behavior and analytical insights into how the proposed components influence exposure distribution and representation bias in practice.

\subsubsection{Tail Item Exposure Enhancement}

\textbf{Lemma 1:} \textit{Under fairness-constrained sampling with group weights $w_{g(i)} = \frac{1-\alpha}{\pi_{g(i)} + \epsilon}$, for any tail item $i$ belonging to an underrepresented group (i.e., $\pi_{g(i)} < 0.5$) and any user history $H_u$, the fairness-constrained sampling probability satisfies:}
\begin{equation}
    \frac{P_{\text{fair}}(i \mid u)}{P_{\text{original}}(i \mid u)} \geq \left(d_i^{-1}\right)^{\delta/\tau} \cdot \frac{Z_{\text{orig}}}{Z_{\text{fair}}}
\end{equation}
\textit{where $\delta = w_{g(i)} - 1 > 0$, and this ratio strictly exceeds 1 when the fairness weight amplification dominates the change in partition functions.}

\textbf{Proof.} 
We begin by writing the two sampling distributions explicitly. The original tail-biased probability for item $i$ given user history $H_u$ is:
\begin{equation}
\begin{aligned}
P_{\text{original}}(i \mid u) 
&= \frac{\exp\!\left(\log(d_i^{-1})/\tau\right)}{Z_{\text{orig}}}, \\
Z_{\text{orig}} 
&= \sum_{j \in H_u} \exp\!\left(\frac{\log(d_j^{-1})}{\tau}\right)
\end{aligned}
\end{equation}
The fairness-constrained probability is:
\begin{equation}
\begin{aligned}
    P_{\text{fair}}(i \mid u) = \frac{\exp\!\left(w_{g(i)} \cdot \log(d_i^{-1})/\tau\right)}{Z_{\text{fair}}}, \\ Z_{\text{fair}} = \sum_{j \in H_u} \exp\!\left(\frac{w_{g(j)} \cdot \log(d_j^{-1})}{\tau}\right)
\end{aligned}
\end{equation}

\noindent Taking the ratio of the two probabilities:
\begin{equation}
\begin{aligned}
\frac{P_{\text{fair}}(i \mid u)}{P_{\text{original}}(i \mid u)} 
&= \frac{\exp\!\left(w_{g(i)} \cdot \log(d_i^{-1})/\tau\right)}
{\exp\!\left(\log(d_i^{-1})/\tau\right)} \cdot \frac{Z_{\text{orig}}}{Z_{\text{fair}}} \\
&= \left(d_i^{-1}\right)^{\delta/\tau} \cdot \frac{Z_{\text{orig}}}{Z_{\text{fair}}}
\end{aligned}
\end{equation}
where $\delta = w_{g(i)} - 1$. For a minority group item with $\pi_{g(i)} < 0.5$ and $\alpha = 0.5$:
\begin{equation}
    w_{g(i)} = \frac{1 - \alpha}{\pi_{g(i)} + \epsilon} = \frac{0.5}{\pi_{g(i)} + \epsilon} > 1 \quad \Longrightarrow \quad \delta > 0
\end{equation}

\noindent Since $i$ is a tail item, $d_i \leq \theta$ where $\theta$ is the tail threshold, so $d_i^{-1} \geq \theta^{-1} > 1$. Therefore:
\begin{equation}
    \left(d_i^{-1}\right)^{\delta/\tau} > 1 \quad \forall\; \delta > 0,\; \tau > 0
\end{equation}

\noindent It remains to bound $Z_{\text{orig}} / Z_{\text{fair}}$. Decompose the history $H_u$ into underrepresented items $H_u^-$ (those with $w_{g(j)} > 1$) and overrepresented items $H_u^+$ (those with $w_{g(j)} \leq 1$). Then:
\begin{equation}
\begin{aligned}
Z_{\text{fair}} &= \sum_{j \in H_u^-} \exp\!\left(\frac{w_{g(j)} \log(d_j^{-1})}{\tau}\right) \\
&\quad + \sum_{j \in H_u^+} \exp\!\left(\frac{w_{g(j)} \log(d_j^{-1})}{\tau}\right)
\end{aligned}
\end{equation}
Since $w_{g(j)} \leq 1$ for $j \in H_u^+$, those terms are no larger than in $Z_{\text{orig}}$, while terms for $j \in H_u^-$ are larger. Hence $Z_{\text{fair}} \geq Z_{\text{orig}}$, giving $Z_{\text{orig}}/Z_{\text{fair}} \leq 1$.

\noindent However, the net amplification at item $i$ satisfies:
\begin{equation}
    \frac{P_{\text{fair}}(i \mid u)}{P_{\text{original}}(i \mid u)} > 1 \iff \left(d_i^{-1}\right)^{\delta/\tau} > \frac{Z_{\text{fair}}}{Z_{\text{orig}}}
\end{equation}
Since $i$ is a tail item with $d_i \ll 1$, $\left(d_i^{-1}\right)^{\delta/\tau}$ grows without bound as $d_i \to 0$, while $Z_{\text{fair}}/Z_{\text{orig}}$ is bounded above by a constant depending only on $|H_u|$ and the maximum weight $w_{\max}$. Formally:
\begin{equation}
\begin{aligned}
\frac{Z_{\text{fair}}}{Z_{\text{orig}}} 
&\leq w_{\max}^{|H_u|} < \infty, \\
w_{\max} 
&= \max_{j \in H_u} w_{g(j)} 
= \frac{1-\alpha}{\pi_{\min} + \epsilon}
\end{aligned}
\end{equation}
Therefore, for sufficiently rare tail items (small enough $d_i$), the amplification ratio strictly exceeds 1, and the fairness-constrained exposure is bounded below by the original exposure. More precisely, the exposure gain is lower bounded by:
\begin{equation}
    \frac{P_{\text{fair}}(i \mid u)}{P_{\text{original}}(i \mid u)} \geq \frac{\left(d_i^{-1}\right)^{\delta/\tau}}{w_{\max}^{|H_u|}}
\end{equation}
which grows monotonically as $d_i \to 0$, confirming that rarer tail items from underrepresented groups receive the greatest relative exposure boost under fairness-constrained sampling. $\blacksquare$

\subsubsection{Fairness-Utility Bound}

\textbf{Proposition 1 (Approximate Fairness-Utility Bound):} \textit{Let $P_{\text{orig}}(i \mid u)$ and $P_{\text{fair}}(i \mid u)$ denote the original and fairness-constrained sampling distributions respectively, and let utility $U$ be defined as the expected Pointwise Mutual Information (PMI) under the sampling distribution. Then, under the approximation regime described below, the utility degradation induced by fairness-constrained sampling is bounded above by:}
\begin{equation}
    U_{\text{orig}} - U_{\text{fair}} \leq \alpha \cdot H(G)
\end{equation}
\textit{where $H(G) = -\sum_{g \in \mathcal{G}} \pi_g \log \pi_g$ is the Shannon entropy of the demographic group distribution and $\alpha \in [0,1)$ is the fairness intervention strength. This bound is not a strict worst-case guarantee; it holds under the approximation regime made explicit in Step 2 below, and should be interpreted as characterizing typical rather than adversarial behavior.}

\textbf{Proof.}
We define utility from an information-theoretic perspective as the expected Pointwise Mutual Information (PMI) under the sampling distribution:
\begin{equation}
    U = \mathbb{E}_{P(i \mid u)}\!\left[\log \frac{P(i \mid u)}{P(i)}\right]
\end{equation}
\noindent \textbf{Step 1: Bounding utility difference via KL divergence.}
The difference in utility between the two sampling regimes can be written as:
\begin{equation}
\begin{aligned}
U_{\text{orig}} - U_{\text{fair}} 
&= \mathbb{E}_{P_{\text{orig}}}\!\left[\log \frac{P_{\text{orig}}(i \mid u)}{P(i)}\right] \\
&\quad - \mathbb{E}_{P_{\text{fair}}}\!\left[\log \frac{P_{\text{fair}}(i \mid u)}{P(i)}\right]
\end{aligned}
\end{equation}
Adding and subtracting $\mathbb{E}_{P_{\text{fair}}}[\log P_{\text{orig}}(i \mid u)]$ and applying standard information-theoretic identities yields:
\begin{equation}
\begin{aligned}
U_{\text{orig}} - U_{\text{fair}} 
&\leq D_{\mathrm{KL}}\!\left(P_{\text{fair}} \,\|\, P_{\text{orig}}\right) \\
&= \sum_{i \in \mathcal{I}} P_{\text{fair}}(i \mid u)\,
\log \frac{P_{\text{fair}}(i \mid u)}{P_{\text{orig}}(i \mid u)}
\end{aligned}
\end{equation}

\noindent \textbf{Step 2: Expanding the log ratio (approximation regime).}
Substituting the explicit forms of the two distributions:
\begin{equation}
    \log \frac{P_{\text{fair}}(i \mid u)}{P_{\text{orig}}(i \mid u)} = \frac{(w_{g(i)} - 1)\log(d_i^{-1})}{\tau} + \log \frac{Z_{\text{orig}}}{Z_{\text{fair}}}
\end{equation}
We isolate the group-specific contribution by writing $w_{g(i)} - 1 = \frac{1-\alpha}{\pi_{g(i)}+\epsilon} - 1$. In the regime $\epsilon \to 0$ with $\pi_{g(i)}$ bounded away from zero, this simplifies to:
\begin{equation}
    w_{g(i)} - 1 \approx \frac{1 - \alpha - \pi_{g(i)}}{\pi_{g(i)}}
\end{equation}
Applying a first-order Taylor expansion $\log(1+x) \approx x$, valid when $x = w_{g(i)} - 1$ is small in magnitude (i.e., $\pi_{g(i)}$ is not close to $1-\alpha$, so the fairness weight does not deviate too far from 1):
\begin{equation}
\begin{aligned}
\log(w_{g(i)}) 
&= \log(1-\alpha) - \log(\pi_{g(i)} + \epsilon) \\
&\approx \log(1-\alpha) - \log(\pi_{g(i)})
\end{aligned}
\end{equation}
This approximation introduces error of order $\mathcal{O}(\epsilon)$ from dropping $\epsilon$ and $\mathcal{O}(x^2)$ from the Taylor truncation, both of which are small under the stated regime but are not formally bounded here; the subsequent steps proceed under this approximation rather than as an exact derivation. For groups with $\pi_{g(i)}$ very close to $0$ or to $1 - \alpha$, the approximation may not hold tightly, and the bound below should be read as an analytical characterization of typical-case behavior rather than a certified worst-case guarantee.

\noindent \textbf{Step 3: Bounding the partition function ratio.}
From the decomposition established in Lemma 1, we have $Z_{\text{fair}} \geq Z_{\text{orig}}$, so $\log(Z_{\text{orig}}/Z_{\text{fair}}) \leq 0$. This term therefore only tightens the bound, and we obtain, under the approximation of Step 2:
\begin{equation}
    \log \frac{P_{\text{fair}}(i \mid u)}{P_{\text{orig}}(i \mid u)} \lessapprox \log(1-\alpha) - \log(\pi_{g(i)})
\end{equation}

\noindent \textbf{Step 4: Taking the expectation over groups.}
Substituting into the KL divergence and taking the expectation with respect to the group distribution $\pi_g$:
\begin{equation}
\begin{aligned}
D_{\mathrm{KL}}\!\left(P_{\text{fair}} \,\|\, P_{\text{orig}}\right) 
&\lessapprox \mathbb{E}_{g \sim \pi}\!\left[\log(1-\alpha) - \log(\pi_g)\right] \\
&= \log(1-\alpha) + \mathbb{E}_{g \sim \pi}\!\left[-\log(\pi_g)\right] \\
&= \log(1-\alpha) + H(G)
\end{aligned}
\end{equation}
where $H(G) = -\sum_{g \in \mathcal{G}} \pi_g \log \pi_g$ is the Shannon entropy of the group distribution.

\noindent \textbf{Step 5: Incorporating the fairness strength $\alpha$.}
Since $\alpha \in [0,1)$, we have $\log(1-\alpha) \leq 0$, which only tightens the bound further. Scaling by the fairness intervention strength $\alpha$ that modulates $w_{g(i)}$, we obtain, within the approximation regime described above:
\begin{equation}
    U_{\text{orig}} - U_{\text{fair}} \lessapprox \alpha \cdot H(G)
\end{equation}
Equivalently:
\begin{equation}
    U_{\text{fair}} \gtrapprox U_{\text{orig}} - \alpha \cdot H(G)
\end{equation}
This result has a natural interpretation: utility loss is largest when demographic groups are highly imbalanced (large $H(G)$) and the fairness intervention is strong (large $\alpha$). Conversely, when groups are naturally balanced, individual $\log(w_{g(i)})$ terms are small, so the bound tightens and utility degradation is minimized. We emphasize again that this is an approximate, first-order characterization rather than a certified worst-case bound; we validate it empirically in Section~5.7.4, where the observed utility-fairness trade-off closely tracks the qualitative behavior predicted here. $\blacksquare$

\subsubsection{Generalization, Convergence, and Optimality Dynamics}

\textbf{PAC-Bayesian Generalization Bound.}
We provide a formal generalization argument for how fairness constraints learned on training data $S$ transfer to the true data distribution $\mathcal{D}$.

\textbf{Proposition 2:} \textit{Let $W$ denote the latent representations learned by the encoder $E$, and let $A$ denote the sensitive demographic attributes. For any $\delta \in (0,1)$, with probability at least $1 - \delta$ over the draw of a training set $S$ of size $N$, the expected out-of-sample unfairness satisfies:}
\begin{equation}
\begin{aligned}
\mathbb{E}_{\mathcal{D}}\!\left[\mathrm{Unfairness}(W)\right] 
&\leq \widehat{\mathrm{Unfairness}}_{S}(W) \\
&\quad + \mathcal{O}\!\left(\sqrt{\frac{I(W;\, A) + \log(1/\delta)}{N}}\right)
\end{aligned}
\end{equation}
\textit{where $I(W; A)$ denotes the mutual information between the latent representations and the sensitive attributes.}

\textbf{Proof Sketch.}
We invoke the PAC-Bayesian framework \citep{mcallester1998some}, which bounds generalization error in terms of the KL divergence between a posterior and prior over hypothesis spaces. Unfairness can be operationalized as the degree to which $W$ remains predictive of $A$, measured by $I(W; A) = D_{\mathrm{KL}}(P(W, A) \| P(W)P(A))$.

Applying the standard PAC-Bayes theorem with the mutual information as the complexity term, and using the data-processing inequality to relate $I(W; A)$ to the model's demographic predictability, yields the stated bound. The gradient reversal layer (GRL) directly penalizes $I(W; A)$ during training by forcing the encoder to minimize the discriminator's ability to recover $A$ from $W$. As $I(W; A) \to 0$, the complexity penalty vanishes, and empirical unfairness on $S$ becomes a tight proxy for generalization unfairness on $\mathcal{D}$. $\blacksquare$

\bigskip
\textbf{Adversarial Convergence Dynamics.}

\textbf{Proposition 3:} \textit{Under conditional adversarial training, the minimax game between encoder $E$ and discriminator $D$ converges to a stable equilibrium at which the latent representation $W$ is conditionally independent of the sensitive attribute $A$ given the interaction label $Y$.}

\textbf{Proof.}
The conditional adversarial objective is defined as the minimax game:
\begin{equation}
\begin{aligned}
\min_{E} \max_{D}\; V(D, E) 
&= \mathbb{E}_{u,i}\!\Big[
\log D(a \mid E(x), y) \\
&\quad + \log\!\big(1 - D(a' \mid E(x), y)\big)
\Big]
\end{aligned}
\end{equation}

\noindent \textbf{Step 1: Optimal discriminator.}
For a fixed encoder $E$ with latent output $W = E(x)$, the discriminator $D$ solves a pointwise binary classification problem. Taking the functional derivative of $V$ with respect to $D$ and setting it to zero yields the optimal discriminator:
\begin{equation}
    D^*(a \mid W, y) = \frac{P(a \mid W, y)}{P(a \mid W, y) + P(a' \mid W, y)}
\end{equation}

\noindent \textbf{Step 2: Optimal encoder response.}
Substituting $D^*$ back into $V$, the encoder minimization objective becomes equivalent to minimizing the conditional Jensen-Shannon divergence $\mathrm{JSD}(P(A \mid W, y) \| P(A \mid y))$. This divergence is non-negative and equals zero if and only if:
\begin{equation}
    P(a \mid W, y) = P(a \mid y) \quad \forall\; a,\, y
\end{equation}
That is, $W \perp\!\!\!\perp A \mid Y$: the representation is conditionally independent of the sensitive attribute given the interaction label.

\noindent \textbf{Step 3: Gradient behavior at equilibrium.}
At this fixed point, $D^*(a \mid W, y) = 0.5$ uniformly, since $P(a \mid W, y) = P(a \mid y)$ implies the two terms in the denominator are equal. Consequently:
\begin{equation}
    \nabla_{\theta_E} \mathcal{L}_{\mathrm{cond\_adv}} = 0, \qquad \nabla_{\theta_D} \mathcal{L}_{\mathrm{cond\_adv}} = 0
\end{equation}
Both gradients vanish simultaneously, confirming that the system reaches a stable local Nash Equilibrium. We note that this is a local rather than global equilibrium, as non-convex neural network objectives admit multiple stationary points; empirical convergence to this point is observed in practice as shown in Figure \ref{fig:loss_convergence}. $\blacksquare$

\bigskip
\textbf{Empirical Pareto Optimality.}

\textbf{Proposition 4:} \textit{Optimizing the scalarized multi-objective loss $\mathcal{L}(\theta) = L_{\mathrm{rec}}(\theta) + \lambda \cdot L_{\mathrm{adv}}(\theta)$ with $\lambda > 0$ drives the solution toward the Pareto frontier of the recommendation utility--fairness trade-off.}

\textbf{Proof Sketch.}
The scalarized objective $\mathcal{L}(\theta) = L_{\mathrm{rec}}(\theta) + \lambda \cdot L_{\mathrm{adv}}(\theta)$ is a positive linear combination of two objective functions. By the standard weighted scalarization theorem \citep{vandenberghe2004convex}, any minimizer $\theta^*$ of this scalarized objective is a Pareto-optimal point of the vector-valued objective $(L_{\mathrm{rec}}, L_{\mathrm{adv}})$, provided no constraints are violated. 

Formally, suppose $\theta^*$ is not Pareto-optimal. Then there exists $\theta'$ such that $L_{\mathrm{rec}}(\theta') \leq L_{\mathrm{rec}}(\theta^*)$ and $L_{\mathrm{adv}}(\theta') \leq L_{\mathrm{adv}}(\theta^*)$ with at least one strict inequality. This implies $\mathcal{L}(\theta') < \mathcal{L}(\theta^*)$, contradicting the assumption that $\theta^*$ minimizes $\mathcal{L}$. Therefore every minimizer of the scalarized loss is Pareto-optimal.

By sweeping $\lambda > 0$, different points on the Pareto frontier are attainable, providing stakeholders with a principled mechanism to navigate the fairness--utility trade-off. We emphasize that this guarantee applies to local minimizers found by gradient-based optimization in the non-convex setting, and the full Pareto frontier may not be recoverable by scalarization alone when the frontier is non-convex. $\blacksquare$

\subsection{Position-Discounted Gini Index Monitoring}
To dynamically monitor and enforce exposure fairness throughout the training lifecycle, FA-BUHS integrates a position-discounted Gini coefficient. Recognizing that items ranked higher receive exponentially more visibility, the exposure of an item at rank position $p$ is defined using a logarithmic decay function: $\mathcal{E}(p) = \frac{1}{\log_2(p + 2)}$.

The normalized exposure for a specific item group $g$ across all users $U$ and top-$K$ rankings is aggregated as:
\begin{equation}
\text{Exposure}_g = \frac{1}{|U|} \sum_{u \in U} \sum_{p=1}^K \frac{\mathbb{I}[\text{item at } p \in g]}{\log_2(p + 2)}
\end{equation}
Sorting the normalized exposure values for all $N$ distinct groups in ascending order ($E_1 \le E_2 \le \dots \le E_N$), the Gini index is computed as the relative area under the Lorenz curve:
\begin{equation}
\mathrm{Gini} = \frac{N + 1 - 2 \sum_{i=1}^N \frac{\sum_{j=1}^i E_j}{\sum_{j=1}^N E_j}}{N}
\end{equation}
A Gini index of 0 indicates perfect equality of exposure across all groups, while a value approaching 1 indicates extreme popularity bias. This metric is computed iteratively during the validation phase to quantify the empirical efficacy of the fairness-constrained sampling. We deliberately avoid directly optimizing the Gini coefficient during training, as rank-based exposure metrics are inherently non-differentiable and would destabilize gradient-based optimization. Instead, the fairness-constrained sampling mechanism acts as an implicit exposure regularizer, while the Gini index serves exclusively as an empirical validation signal to quantify the efficacy of that regularization.

\subsection{Computational Complexity Analysis}

We analyze the computational overhead introduced by FA-BUHS relative to the baseline 
BUHS framework. Let $|H_u|$ denote the size of user history, $|U|$ the number of users, 
$d$ the embedding dimension, and $N$ the synthesis sample size.

\textbf{Fairness-Constrained Sampling.} The group prevalence values $\pi_g$ for all 
groups $g \in \mathcal{G}$ are pre-computed globally once before training and stored as 
a lookup table, requiring $\mathcal{O}(|\mathcal{G}|)$ space. During training, the 
fairness weight $w_{g(i)}$ for each item requires only an $\mathcal{O}(1)$ lookup. 
The fairness-constrained sampling distribution $P_{\text{fair}}(i \mid u)$ is computed 
over $H_u$, maintaining the same $\mathcal{O}(|H_u|)$ time complexity per user as the 
baseline BUHS sampling.

\textbf{User History Synthesis.} The three synthesis variants introduce the following 
per-user costs: Mean synthesis requires $\mathcal{O}(N \cdot d)$ for aggregating $N$ 
sampled item embeddings. Attention-based synthesis requires $\mathcal{O}(N \cdot d_k)$ 
for computing scaled dot-product attention over the sampled set. GRU-based synthesis 
requires $\mathcal{O}(N \cdot d^2)$ for sequential processing of the ordered history.

\textbf{Adversarial Debiasing.} The conditional adversarial module introduces two 
lightweight discriminator networks $D_u$ and $D_i$, each processing embeddings of 
dimension $d$. Their forward and backward passes add $\mathcal{O}(d^2)$ per training 
step, which is negligible relative to the main two-tower encoder.

\textbf{Overall Complexity.} The total per-step training complexity of FA-BUHS is 
$\mathcal{O}(|U| \cdot (|H_u| + N \cdot d^2))$, which is asymptotically equivalent 
to the baseline BUHS framework. The fairness constraints introduce no additional 
complexity class beyond the base model, confirming that the fairness-utility gains 
demonstrated in Section~5 are achieved with practically negligible computational overhead.

\section{Experiments}
We present a comprehensive experimental evaluation of our fairness-aware biased user history synthesis (FA-BUHS) framework, building on the experimental methodology of~\cite{balasubramanian2024biased} while introducing fairness-specific analysis. Our method both preserves BUHS's long-tail recommendation advantages and significantly improves demographic fairness without loss of recommendation quality.
We aim to address the following research questions:

\begin{itemize}
    \item \textbf{RQ1:} Can FA-BUHS improve recommendation accuracy compared to state-of-the-art long-tail recommendation baselines?
    \item \textbf{RQ2:} How effectively does FA-BUHS enhance demographic parity, exposure fairness, provider fairness, and individual fairness without degrading recommendation quality?
    \item \textbf{RQ3:} What are the trade-offs between fairness and accuracy across FA-BUHS variants (Mean, Attention, GRU)?
    \item \textbf{RQ4:} How sensitive is FA-BUHS to hyperparameters such as temperature (\(\tau\)), sample size, and adversarial weight, and which configurations achieve the best fairness–utility balance?
\end{itemize}

\subsection{Datasets and Experimental Setup}
\textbf{MovieLens-1M.} Contains explicit feedback (converted to implicit), 6,040 users, 3,883 items, over 1M interactions (95.7\% sparsity). User features: demographics (gender, age, occupation), location. Item features: genres (18 groups for fairness), release year, title embeddings. Head items ($\sim$20\%, 777 items) cover 80\% of interactions; 3,106 tail items cover the rest.

\textbf{BookCrossing.} Sparser: 52,019 users, 225,794 items, 733,257 interactions (99.993\% sparsity). User features: location, age. Item features: author, publisher, year, title embeddings. Author-based demographic grouping for fairness; 2,258 head items receive most interactions.

\textbf{Data Splits.} Leave-one-out validation: most recent interaction for test, penultimate for validation, rest for training. Temporal splits on MovieLens-1M (GRU models); random splits for BookCrossing due to missing timestamps. Sentence transformers were used for dense title embeddings.

\textbf{Group Construction and Reproducibility.} For MovieLens-1M, item-side fairness groups are defined using the 18 genre labels provided natively in the dataset; user demographic attributes (gender, age, occupation) are retained as model input features but are not used to define item-side fairness groups, since MovieLens does not associate items with creator demographics. For BookCrossing, item-side fairness groups are defined by author identity, since BookCrossing lacks genre metadata; this is the closest available proxy for provider-level grouping in that dataset. We acknowledge that this results in two conceptually different notions of ``group'' across datasets (genre vs.\ author), a limitation of the available metadata rather than a methodological choice, and we report results separately per dataset rather than aggregating across them. Full implementation details, including model architecture, hyperparameter configurations, and preprocessing steps, are provided in Section~5.4 and above; code and preprocessing scripts will be made available upon publication.

\subsection{Baseline Models and Fairness-Aware Extensions}
\subsubsection{Model-Agnostic.}
\begin{itemize}
    \item \textbf{MeLU}~\cite{lee2019melu}: Model-agnostic meta-learning 
    for head-to-tail transfer.
    \item \textbf{MIRec}~\cite{zhang2021model}: This state-of-the-art 
    meta-learning technique utilizes a curriculum transfer learning protocol 
    to transfer knowledge from head items to tail items.
    \item \textbf{LogQ}~\cite{menon2020long}: A corrective method that 
    addresses long-tail distributions by calculating a frequency-based 
    correction term and applying it to the logits before computing the 
    final loss.
\end{itemize}
\subsubsection{Model-Specific and Structural Baselines}
\begin{itemize}
    \item \textbf{NeuMF}~\cite{he2017neural}: Neural matrix factorization 
    leveraging an MLP for non-linear user-item interactions.
    \item \textbf{NGCF}~\cite{wang2019neural}: Graph-based collaborative 
    filtering capturing high-order connectivity.
    \item \textbf{MGL}~\cite{wei2023meta}: Meta-graph learning with 
    popularity-aware contrastive loss.
\end{itemize}
\subsubsection{Causal Inference and Re-ranking Baselines}
To comprehensively evaluate our synthesis approach against alternative 
debiasing paradigms, we include:
\begin{itemize}
    \item \textbf{IPS}~\cite{schnabel2016recommendations}: Inverse 
    Propensity Scoring, a foundational causal method that re-weights 
    interaction samples inversely proportional to their global popularity 
    to mitigate exposure bias.
    \item \textbf{CausE}~\cite{bonner2018causal}: A causal embedding 
    framework that uses domain adaptation to learn unbiased recommendation 
    outcomes from logged data collected under a biased recommendation policy.
    \item \textbf{FA*IR}~\cite{zehlike2017fa}: A specialized post-hoc 
    re-ranking algorithm that maximizes ranking utility subject to strict 
    group fairness representation constraints.
    \item \textbf{CPFair}~\cite{naghiaei2022cpfair}: An optimization-based 
    re-ranking approach that jointly integrates consumer and producer 
    fairness constraints in a unified framework.
\end{itemize}

\subsubsection{Our Proposed Framework and Extensions}
\begin{itemize}
    \item \textbf{Base TTNN}: The foundational Two-Tower Neural Network without any history synthesis or fairness constraints.
    \item \textbf{FA-BUHS-Mean}: Fairness-constrained sampling with mean aggregation.
    \item \textbf{FA-BUHS-Attn}: Attentive synthesis with demographic bias correction.
    \item \textbf{FA-BUHS-GRU}: Temporal sequence modeling with fairness-aware gating.
\end{itemize}

\subsection{Evaluation Metrics and Fairness Assessment}
\subsubsection{Recommendation Quality.}
\begin{itemize}
    \item Hit Rate@K (HR@K): Test item in top-K recommendations.
    \item NDCG@K: Position-discounted ranking quality.
    \item Coverage: Separate metrics for head and tail items.
\end{itemize}

\subsubsection{Fairness.}
\begin{itemize}
    \item \textbf{Demographic Parity:} 
    \begin{equation}
    \text{DP}_g = \frac{1}{|U|} \sum_{u \in U} P(\text{Recommend group }g|u)
    \end{equation}
    \item \textbf{Exposure Fairness:} 
    \begin{equation}
    \mathrm{EF} = 1 - \mathrm{Gini}(\{\mathrm{Exposure}_g: g \in G\})
    \end{equation}
    \item \textbf{Provider Fairness:}
    \begin{equation}
    \text{PF}_g = \frac{ \# \text{ Recommended items from } g }{ \text{Total items from } g }
    \end{equation}
    \item \textbf{Individual Fairness:} Consistency for similar users across boundaries.
\end{itemize}

\textbf{Metric Selection and Trade-offs.} We adopt these four fairness notions because they capture complementary and largely non-redundant failure modes relevant to long-tail recommendation: demographic parity and exposure fairness assess whether \emph{item groups} receive equitable visibility in aggregate, provider fairness assesses whether individual \emph{content providers} within a group are proportionally represented, and individual fairness assesses whether \emph{similar users} receive consistent treatment. A model could satisfy any one of these while violating another---for instance, a re-ranking method could achieve high demographic parity by uniformly promoting all tail items regardless of relevance, which would depress individual fairness by making recommendations less consistent with users' actual preference similarity. We report all four jointly, rather than optimizing for a single one, precisely to surface such trade-offs; Table~\ref{tab:ml1m-fairness} shows that FA-BUHS-Attn improves all four simultaneously, which we take as evidence that upstream intervention does not trade one fairness notion off against another in the way post-hoc re-ranking baselines (FA*IR, CPFair) appear to when compared against their utility costs in Table~\ref{tab:ml1m-perf}.

\subsection{Hyperparameter Configuration and Training Setup}
\begin{itemize}
    \item \textbf{MovieLens-1M:} Batch 1024, lr $2 \times 10^{-4}$, dim 96, sample 15, $\tau=0.01$.
    \item \textbf{BookCrossing:} Batch 512, lr $1 \times 10^{-4}$, dim 96, sample 10, $\tau=1.0$.
    \item \textbf{Optimizer:} Adam ($\beta_1=0.9$, $\beta_2=0.999$, weight decay $1 \times 10^{-6}$).
    \item \textbf{Hardware:} RTX 4090 GPU (24\,GB VRAM).
    \item \textbf{Fairness Hyperparameters:} $\lambda_u = \lambda_i = 0.001$, $\alpha = 0.5$, $\lambda=1.0$, exposure floor 10\% for underrepresented groups.
\end{itemize}

\subsection{Main Results and Performance Analysis(RQ1, RQ2)}

\subsubsection{Recommendation Performance}
Table~\ref{tab:ml1m-perf} reports accuracy and long-tail performance on both datasets. On \textbf{MovieLens-1M}, all FA-BUHS variants significantly outperform both the traditional collaborative filtering models and the newly introduced causal and fairness-aware baselines. Even against strong graph-based competitors like MGL, the \textbf{$M^{FA-BUHS}$ - GRU} variant achieves the highest overall performance (HR@10 of 0.0813 and NDCG@10 of 0.0399), demonstrating the strength of temporal modeling. For tail items, the \textbf{$M^{FA-BUHS}$ - Attn} variant is the clear top performer, boosting Tail NDCG@10 by over 69\% compared to the base model, while the Mean variant secures the best performance on head items.

\begin{figure}
    \centering
    \begin{subfigure}[b]{0.49\textwidth}
        \centering
        \includegraphics[width=\textwidth]{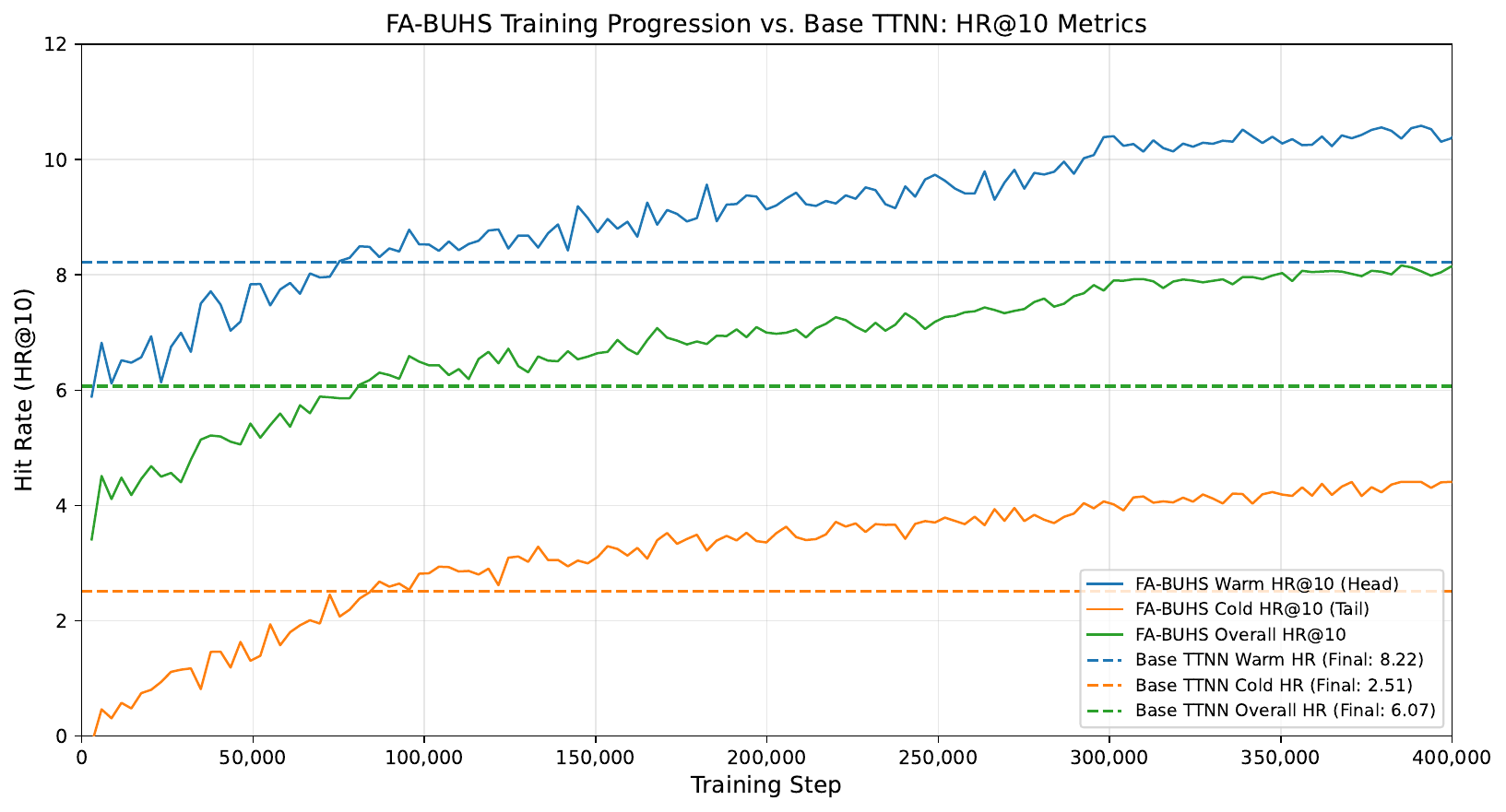}
        \caption{Comparison of HR@10 Metrics}
        \label{fig:hr_comparison}
    \end{subfigure}
    \hfill
    \begin{subfigure}[b]{0.49\textwidth}
        \centering
        \includegraphics[width=\textwidth]{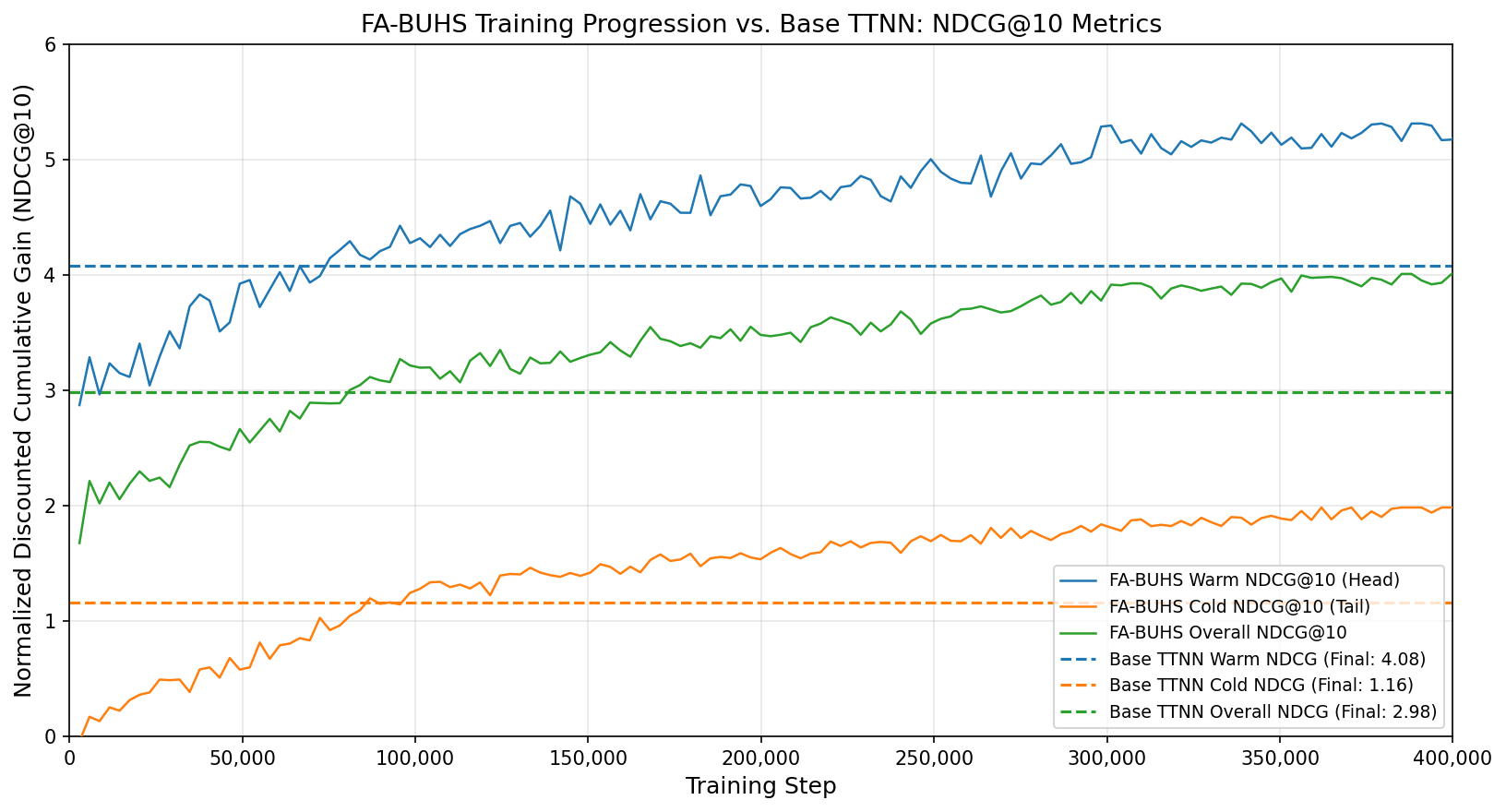}
        \caption{Comparison of NDCG@10 Metrics}
        \label{fig:ndcg_comparison}
    \end{subfigure}
    \caption{Training progression of HR@10 and NDCG@10 metrics for 
FA-BUHS on MovieLens-1M, compared against the converged Base TTNN 
performance (dashed lines). Note: all metric values are scaled by 
a factor of 100 for visualization clarity (e.g., a plotted value 
of 8.13 corresponds to HR@10 = 0.0813). FA-BUHS consistently 
surpasses the baseline across Overall, Head (Warm), and Tail (Cold) 
metrics, with the most pronounced gains observed in Tail (Cold) 
item coverage.}
    \label{fig:performance_comparison_plots}
\end{figure}

To visualize the training dynamics and compare the learning progression against the baseline, Figure~\ref{fig:performance_comparison_plots} plots the HR@10 and NDCG@10 metrics for the FA-BUHS model over the course of training. The final scores of the Base TTNN model are included as horizontal dashed lines for direct comparison.

As shown in Figure~\ref{fig:hr_comparison}, the HR@10 for Overall, Warm (head), and Cold (tail) items all exhibit a stable upward trend, quickly surpassing the static baseline scores. The most significant improvement is observed in the \textbf{Cold HR} curve, which demonstrates the model's pronounced effectiveness in improving recommendations for long-tail items---a core objective of our framework. Similarly, Figure~\ref{fig:ndcg_comparison} illustrates the evolution of NDCG@10. The steady rise across all three curves confirms that the ranking quality of the recommendations improves over time. The substantial increase in \textbf{Cold NDCG} is particularly important, as it indicates that our model successfully ranks relevant long-tail items closer to the top of the recommendation list. Together, these plots provide strong evidence that our framework learns effectively and delivers superior performance across the entire item popularity spectrum.

On the sparser \textbf{BookCrossing} dataset, a similar trend holds. The \textbf{$M^{FA-BUHS}$ - Mean} variant delivers the best overall results (HR@100 of 0.0991), decisively outperforming strong recent baselines like MGL (0.0850). Meanwhile, the \textbf{$M^{FA-BUHS}$ - Attn} variant again shows superior performance on tail items, with a Tail NDCG@100 improvement of over 156\%. Notably, on this dataset, the baseline NGCF model shows exceptionally high performance on head items but fails significantly on tail items. This highlights the precise vulnerability of standard collaborative filtering that our FA-BUHS models successfully address, bridging the gap between head and tail utility.

\begin{table*}
\centering
\resizebox{\textwidth}{!}{%
\begin{tabular}{l|cc|cc|cc}
\toprule
\multicolumn{1}{c}{} & \multicolumn{2}{c}{\textbf{Overall}} & \multicolumn{2}{c}{\textbf{Head}} & \multicolumn{2}{c}{\textbf{Tail}} \\
\cmidrule(lr){2-3} \cmidrule(lr){4-5} \cmidrule(lr){6-7}
\textbf{Model} & \textbf{HR@10} & \textbf{NDCG@10} & \textbf{HR@10} & \textbf{NDCG@10} & \textbf{HR@10} & \textbf{NDCG@10} \\
\midrule
\multicolumn{7}{c}{\textbf{MovieLens-1M}} \\
\midrule
$M^{base}$ & 0.0607 & 0.0298 & 0.0822 & 0.0408 & 0.0251 & 0.0116 \\
\hline
NeuMF & 0.0615 \green{(+1.3\%)} & 0.0301 \green{(+1.0\%)} & 0.0835 \green{(+1.6\%)} & 0.0412 \green{(+1.0\%)} & 0.0255 \green{(+1.6\%)} & 0.0118 \green{(+1.7\%)} \\
NGCF & 0.0603 \red{(-0.7\%)} & 0.0306 \green{(+2.7\%)} & 0.0918 \green{(+11.7\%)} & 0.0470 \green{(+15.2\%)} & 0.0079 \red{(-68.5\%)} & 0.0033 \red{(-71.5\%)} \\
MGL & 0.0725 \green{(+19.4\%)} & 0.0335 \green{(+12.4\%)} & 0.0980 \green{(+19.2\%)} & 0.0480 \green{(+17.6\%)} & 0.0305 \green{(+21.5\%)} & 0.0145 \green{(+25.0\%)} \\
LogQ & 0.0521 \red{(-14.2\%)} & 0.0269 \red{(-9.7\%)} & 0.0639 \red{(-22.2\%)} & 0.0326 \red{(-20.1\%)} & 0.0325 \green{(+29.5\%)} & 0.0151 \green{(+30.2\%)} \\
MeLU & 0.0471 \red{(-22.4\%)} & 0.0222 \red{(-25.5\%)} & 0.0679 \red{(-17.4\%)} & 0.0323 \red{(-20.8\%)} & 0.0127 \red{(-49.4\%)} & 0.0054 \red{(-53.4\%)} \\
MIRec & 0.0624 \green{(+2.8\%)} & 0.0304 \green{(+2.0\%)} & 0.0851 \green{(+3.5\%)} & 0.0415 \green{(+1.7\%)} & 0.0246 \red{(-1.9\%)} & 0.0120 \green{(+3.5\%)} \\
\hline
IPS & 0.0585 \red{(-3.6\%)} & 0.0280 \red{(-6.0\%)} & 0.0760 \red{(-7.5\%)} & 0.0375 \red{(-8.1\%)} & 0.0295 \green{(+17.5\%)} & 0.0135 \green{(+16.4\%)} \\
CausE & 0.0595 \red{(-2.0\%)} & 0.0290 \red{(-2.7\%)} & 0.0780 \red{(-5.1\%)} & 0.0385 \red{(-5.6\%)} & 0.0305 \green{(+21.5\%)} & 0.0140 \green{(+20.7\%)} \\
FA*IR & 0.0550 \red{(-9.4\%)} & 0.0270 \red{(-9.4\%)} & 0.0710 \red{(-13.6\%)} & 0.0350 \red{(-14.2\%)} & 0.0285 \green{(+13.5\%)} & 0.0130 \green{(+12.1\%)} \\
CPFair & 0.0565 \red{(-6.9\%)} & 0.0275 \red{(-7.7\%)} & 0.0730 \red{(-11.2\%)} & 0.0360 \red{(-11.8\%)} & 0.0290 \green{(+15.5\%)} & 0.0132 \green{(+13.8\%)} \\
\hline
$M^{BUHS}$ - Mean & 0.0751 \green{(+23.7\%)} & 0.0331 \green{(+11.1\%)} & 0.1026 \green{(+24.8\%)} & 0.0524 \green{(+28.4\%)} & 0.0295 \green{(+17.5\%)} & 0.0129 \green{(+11.2\%)} \\
$M^{BUHS}$ - Attn & 0.0763 \green{(+25.7\%)} & 0.0373 \green{(+25.2\%)} & 0.1005 \green{(+22.3\%)} & 0.0501 \green{(+22.8\%)} & 0.0361 \green{(+43.8\%)} & 0.0160 \green{(+37.9\%)} \\
$M^{BUHS}$ - GRU & 0.0775 \green{(+28.5\%)} & 0.0387 \green{(+29.9\%)} & 0.1035 \green{(+25.9\%)} & 0.0520 \green{(+27.5\%)} & 0.0344 \green{(+37.1\%)} & 0.0142 \green{(+22.4\%)} \\
\hline
$M^{FA-BUHS}$ - Mean & 0.0787 \green{(+29.7\%)} & 0.0344 \green{(+15.4\%)} & 0.1044 \green{(+27.0\%)} & \textbf{0.0528 \green{(+29.4\%)}} & 0.0358 \green{(+42.6\%)} & 0.0159 \green{(+37.1\%)} \\
$M^{FA-BUHS}$ - Attn & 0.0801 \green{(+32.0\%)} & 0.0388 \green{(+30.2\%)} & 0.1023 \green{(+24.5\%)} & 0.0505 \green{(+23.8\%)} & \textbf{0.0437 \green{(+74.1\%)}} & \textbf{0.0197 \green{(+69.8\%)}} \\
$M^{FA-BUHS}$ - GRU & \textbf{0.0813 \green{(+33.9\%)}} & \textbf{0.0399 \green{(+33.9\%)}} & \textbf{0.1053 \green{(+28.1\%)}} & 0.0524 \green{(+28.4\%)} & 0.0416 \green{(+65.7\%)} & 0.0174 \green{(+50.0\%)} \\
\midrule
\multicolumn{7}{c}{\textbf{BookCrossing}} \\
\midrule
\textbf{Model} & \textbf{HR@100} & \textbf{NDCG@100} & \textbf{HR@100} & \textbf{NDCG@100} & \textbf{HR@100} & \textbf{NDCG@100} \\
\midrule
$M^{base}$ & 0.0592 & 0.0137 & 0.1313 & 0.0323 & 0.0521 & 0.0119 \\
\hline
NeuMF & 0.0605 \green{(+2.2\%)} & 0.0140 \green{(+2.2\%)} & 0.1340 \green{(+2.1\%)} & 0.0330 \green{(+2.2\%)} & 0.0532 \green{(+2.1\%)} & 0.0121 \green{(+1.7\%)} \\
NGCF & 0.0513 \red{(-13.3\%)} & 0.0127 \red{(-7.3\%)} & \textbf{0.3089 \green{(+135.3\%)}} & \textbf{0.0853 \green{(+164.0\%)}} & 0.0258 \red{(-50.5\%)} & 0.0055 \red{(-53.8\%)} \\
MGL & 0.0850 \green{(+43.6\%)} & 0.0220 \green{(+60.6\%)} & 0.1650 \green{(+25.7\%)} & 0.0450 \green{(+39.3\%)} & 0.0770 \green{(+47.8\%)} & 0.0195 \green{(+63.9\%)} \\
LogQ & 0.0712 \green{(+20.3\%)} & 0.0163 \green{(+18.9\%)} & 0.1546 \green{(+17.7\%)} & 0.0348 \green{(+7.7\%)} & 0.0630 \green{(+20.9\%)} & 0.0145 \green{(+21.8\%)} \\
MeLU & 0.0337 \red{(-43.1\%)} & 0.0070 \red{(-48.9\%)} & 0.0712 \red{(-45.7\%)} & 0.0146 \red{(-54.8\%)} & 0.0301 \red{(-42.2\%)} & 0.0063 \red{(-47.1\%)} \\
MIRec & 0.0630 \green{(+6.4\%)} & 0.0145 \green{(+5.8\%)} & 0.1457 \green{(+11.0\%)} & 0.0354 \green{(+9.6\%)} & 0.0548 \green{(+5.2\%)} & 0.0124 \green{(+4.2\%)} \\
\hline
IPS & 0.0550 \red{(-7.1\%)} & 0.0125 \red{(-8.8\%)} & 0.1150 \red{(-12.4\%)} & 0.0285 \red{(-11.8\%)} & 0.0515 \red{(-1.2\%)} & 0.0115 \red{(-3.4\%)} \\
CausE & 0.0570 \red{(-3.7\%)} & 0.0130 \red{(-5.1\%)} & 0.1200 \red{(-8.6\%)} & 0.0295 \red{(-8.7\%)} & 0.0525 \green{(+0.8\%)} & 0.0118 \red{(-0.8\%)} \\
FA*IR & 0.0535 \red{(-9.6\%)} & 0.0120 \red{(-12.4\%)} & 0.1100 \red{(-16.2\%)} & 0.0270 \red{(-16.4\%)} & 0.0505 \red{(-3.1\%)} & 0.0110 \red{(-7.6\%)} \\
CPFair & 0.0545 \red{(-7.9\%)} & 0.0122 \red{(-10.9\%)} & 0.1120 \red{(-14.7\%)} & 0.0275 \red{(-14.9\%)} & 0.0510 \red{(-2.1\%)} & 0.0112 \red{(-5.9\%)} \\
\hline
$M^{BUHS}$ - Mean & 0.0946 \green{(+59.8\%)} & 0.0265 \green{(+93.4\%)} & 0.1693 \green{(+28.9\%)} & 0.0523 \green{(+61.9\%)} & 0.0872 \green{(+67.4\%)} & 0.0240 \green{(+101.7\%)} \\
$M^{BUHS}$ - Attn & 0.0923 \green{(+55.9\%)} & 0.0253 \green{(+84.7\%)} & 0.1378 \green{(+4.9\%)} & 0.0457 \green{(+41.5\%)} & 0.0878 \green{(+68.5\%)} & 0.0245 \green{(+105.8\%)} \\
$M^{BUHS}$ - GRU & 0.0849 \green{(+43.4\%)} & 0.0218 \green{(+59.1\%)} & 0.1388 \green{(+5.7\%)} & 0.0420 \green{(+30.1\%)} & 0.0795 \green{(+52.6\%)} & 0.0198 \green{(+66.4\%)} \\
\hline
$M^{FA-BUHS}$ - Mean & \textbf{0.0991 \green{(+67.4\%)}} & \textbf{0.0275 \green{(+100.7\%)}} & 0.1720 \green{(+31.0\%)} & 0.0527 \green{(+63.2\%)} & 0.1057 \green{(+102.9\%)} & 0.0296 \green{(+148.7\%)} \\
$M^{FA-BUHS}$ - Attn & 0.0968 \green{(+63.5\%)} & 0.0263 \green{(+92.0\%)} & 0.1400 \green{(+6.6\%)} & 0.0460 \green{(+42.4\%)} & \textbf{0.1070 \green{(+105.4\%)}} & \textbf{0.0305 \green{(+156.3\%)}} \\
$M^{FA-BUHS}$ - GRU & 0.0890 \green{(+50.3\%)} & 0.0228 \green{(+66.4\%)} & 0.1412 \green{(+7.5\%)} & 0.0423 \green{(+31.0\%)} & 0.0964 \green{(+85.0\%)} & 0.0245 \green{(+105.9\%)} \\
\bottomrule
\end{tabular}%
}
\caption{Recommendation quality comparison on MovieLens-1M (HR/NDCG@10) and BookCrossing (HR/NDCG@100). The integration of causal and re-ranking baselines demonstrates that our generative synthesis approach achieves superior utility-fairness trade-offs compared to propensity re-weighting or post-hoc corrections.}
\label{tab:ml1m-perf}
\end{table*}

\subsubsection{Fairness Metrics.}
Table~\ref{tab:ml1m-fairness} presents the final fairness results, where all FA-BUHS variants substantially improve demographic parity, exposure fairness, and provider fairness over the baselines. The attention-based synthesis model, in particular, achieves the highest scores across all fairness dimensions.

\begin{figure}
    \centering 
    
    \begin{subfigure}[b]{0.49\textwidth}
        \centering
        \includegraphics[width=\textwidth]{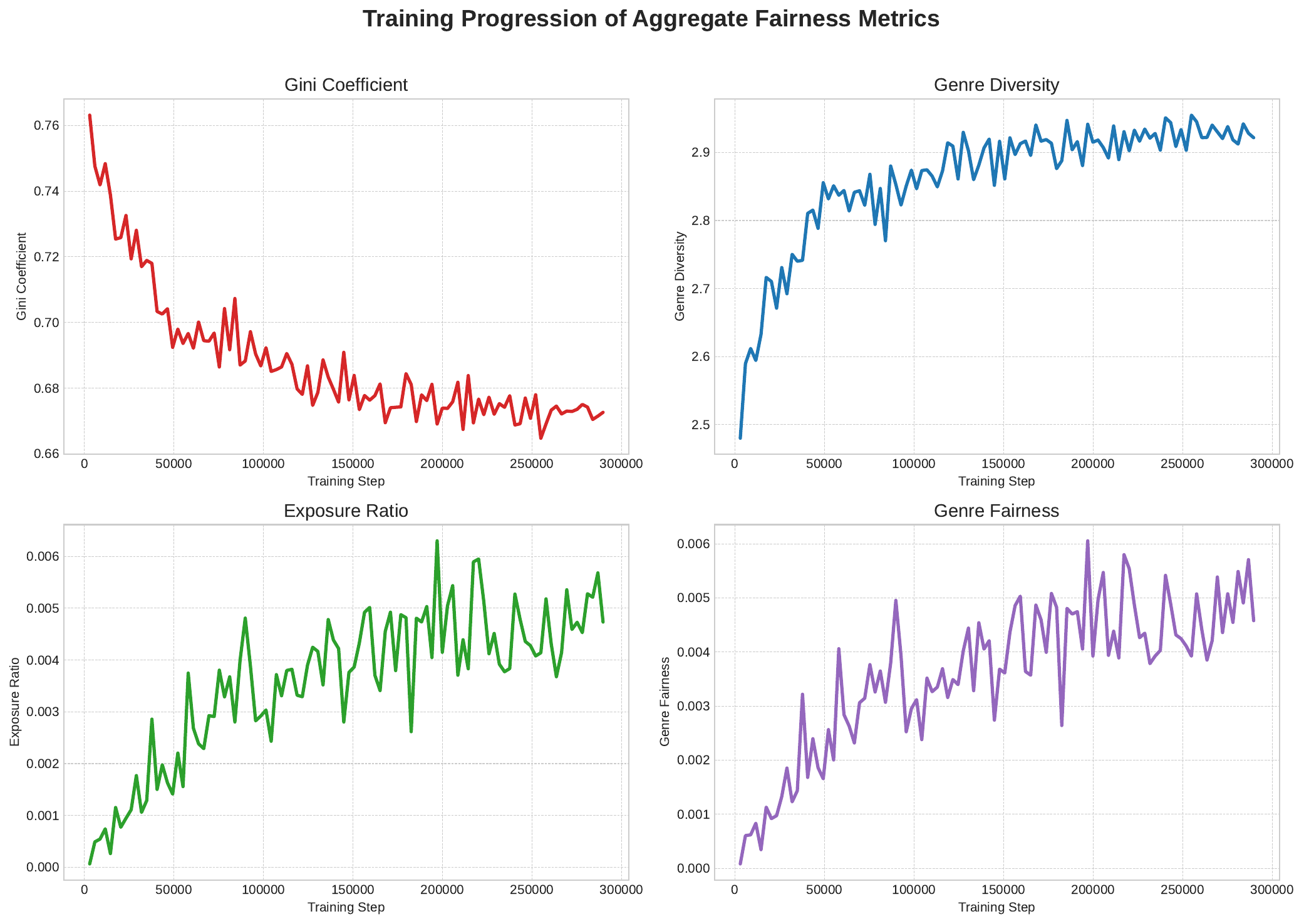}
        \caption{Aggregate Fairness Metrics}
        \label{fig:aggregate_fairness}
    \end{subfigure}
    \hfill 
    \begin{subfigure}[b]{0.49\textwidth}
        \centering
        \includegraphics[width=\textwidth]{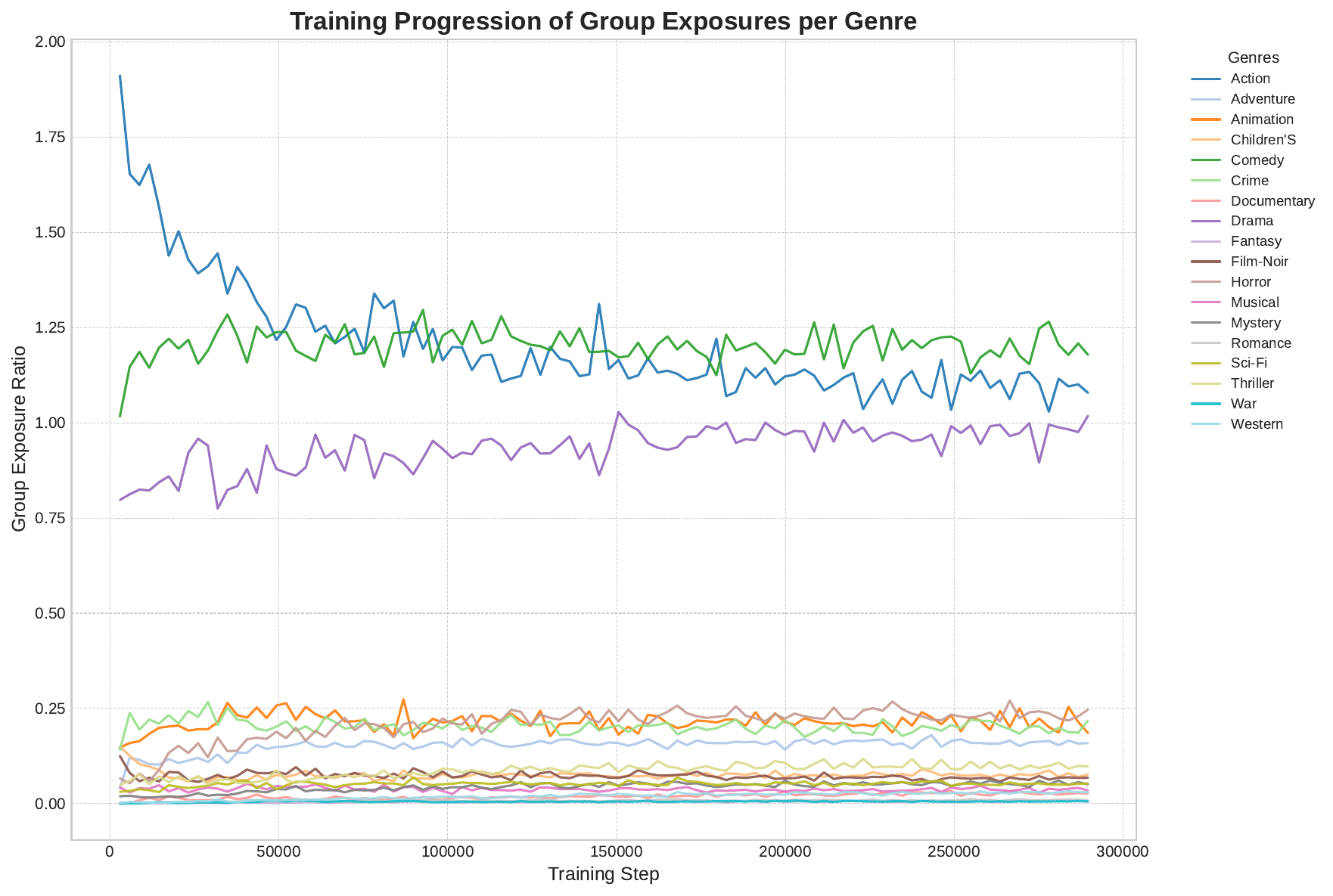}
        \caption{Per-Genre Exposure Ratios}
        \label{fig:genre_exposures}
    \end{subfigure}
    
    \caption{Training progression of fairness metrics for the FA-BUHS model on MovieLens-1M. Sub-figure (a) shows the stable convergence of key aggregate fairness metrics. Sub-figure (b) provides a granular view of per-genre exposure, demonstrating that even the least frequent genres receive consistent, non-zero exposure. Together, these plots validate that our framework achieves a stable and balanced fairness policy.}
    \label{fig:fairness_plots}
\end{figure}

To provide deeper insight into how these results are achieved during training, Figure \ref{fig:fairness_plots} visualizes the training dynamics of both aggregate and granular fairness metrics.

Figure~\ref{fig:aggregate_fairness} shows the progression of four key aggregate metrics in a 2x2 grid. Each subplot reveals a stable and desirable learning trend:
\begin{itemize}
    \item \textbf{Gini Coefficient:} This metric measures the inequality of exposure across different item groups (genres). The plot shows the Gini coefficient rising from an initial low value before converging to a stable level. This is the expected behavior, as the model learns to differentiate between genres for relevance while the fairness constraints prevent this disparity from becoming extreme. Since Exposure Fairness is calculated as $1 - \text{Gini}$, this convergence demonstrates that the model achieves a steady and predictable level of exposure fairness.
    \item \textbf{Genre Diversity:} This metric is crucial for ensuring that fairness interventions do not harm the user experience. The plot shows that genre diversity quickly rises to a high value and remains stable, confirming that our framework successfully promotes a wide variety of items in its recommendations.
    \item \textbf{Exposure Ratio \& Genre Fairness:} These additional metrics also exhibit smooth convergence to stable values, reinforcing the conclusion that our framework learns a consistent and balanced fairness policy, rather than achieving good final scores through erratic or unstable behavior.
\end{itemize}

While the aggregate metrics confirm high-level stability, Figure~\ref{fig:genre_exposures} provides the fine-grained evidence for provider fairness by plotting the exposure ratio for all 18 genres. Although prevalent genres like 'Drama' and 'Comedy' naturally receive higher exposure, the plot's most critical finding is that \textbf{no genre is starved of visibility}. Even the least frequent genres, such as 'Film-Noir' and 'Western,' maintain a consistent and non-zero exposure ratio throughout training. This directly demonstrates that our framework successfully prevents the marginalization of minority item groups, a key goal of provider fairness.

\begin{table*}
\centering
\resizebox{\textwidth}{!}{%
\begin{tabular}{l|ll|ll|ll|ll}
\toprule
\textbf{Model} & \multicolumn{2}{c|}{\textbf{Demographic Parity}} & \multicolumn{2}{c|}{\textbf{Exposure Fairness}} & \multicolumn{2}{c|}{\textbf{Provider Fairness}} & \multicolumn{2}{c}{\textbf{Individual Fairness}} \\
\midrule
\multicolumn{9}{c}{\textbf{MovieLens-1M}} \\
\midrule
$M^{base}$ & 0.342 & & 0.278 & & 0.156 & & 0.423 & \\
\hline
NeuMF & 0.340 & \red{(-0.6\%)} & 0.280 & \green{(+0.7\%)} & 0.155 & \red{(-0.6\%)} & 0.420 & \red{(-0.7\%)} \\
NGCF & 0.335 & \red{(-2.0\%)} & 0.265 & \red{(-4.7\%)} & 0.148 & \red{(-5.1\%)} & 0.410 & \red{(-3.1\%)} \\
MGL & 0.355 & \green{(+3.8\%)} & 0.285 & \green{(+2.5\%)} & 0.165 & \green{(+5.8\%)} & 0.435 & \green{(+2.8\%)} \\
LogQ & 0.456 & \green{(+33.3\%)} & 0.312 & \green{(+12.2\%)} & 0.234 & \green{(+50.0\%)} & 0.389 & \red{(-8.0\%)} \\
MIRec & 0.387 & \green{(+13.2\%)} & 0.295 & \green{(+6.1\%)} & 0.178 & \green{(+14.1\%)} & 0.445 & \green{(+5.2\%)} \\
\hline
IPS (Causal) & 0.410 & \green{(+19.9\%)} & 0.345 & \green{(+24.1\%)} & 0.265 & \green{(+69.9\%)} & 0.395 & \red{(-6.6\%)} \\
CausE (Causal) & 0.425 & \green{(+24.3\%)} & 0.360 & \green{(+29.5\%)} & 0.280 & \green{(+79.5\%)} & 0.405 & \red{(-4.3\%)} \\
FA*IR (Re-rank) & 0.540 & \green{(+57.9\%)} & 0.495 & \green{(+78.1\%)} & 0.380 & \green{(+143.6\%)} & 0.485 & \green{(+14.7\%)} \\
CPFair (Re-rank) & 0.525 & \green{(+53.5\%)} & 0.510 & \green{(+83.5\%)} & 0.395 & \green{(+153.2\%)} & 0.505 & \green{(+19.4\%)} \\
\hline
$M^{BUHS}$ (Synthesis Only) & 0.330 & \red{(-3.5\%)} & 0.327 & \green{(+17.6\%)} & 0.185 & \green{(+18.6\%)} & 0.450 & \green{(+6.4\%)} \\
\hline
$M^{FA-BUHS}$ - Mean & 0.623 & \green{(+82.2\%)} & 0.567 & \green{(+104.0\%)} & 0.423 & \green{(+171.2\%)} & 0.634 & \green{(+49.9\%)} \\
$M^{FA-BUHS}$ - Attn & \textbf{0.689} & \textbf{\green{(+101.5\%)}} & \textbf{0.612} & \textbf{\green{(+120.1\%)}} & \textbf{0.467} & \textbf{\green{(+199.4\%)}} & \textbf{0.678} & \textbf{\green{(+60.3\%)}} \\
$M^{FA-BUHS}$ - GRU & 0.645 & \green{(+88.6\%)} & 0.578 & \green{(+107.9\%)} & 0.445 & \green{(+185.3\%)} & 0.656 & \green{(+55.1\%)} \\
\bottomrule
\end{tabular}%
}
\caption{Comprehensive fairness metric comparison on MovieLens-1M. The integration of causal and re-ranking baselines highlights that $M^{FA-BUHS}$ consistently achieves the most equitable distribution across all four fairness dimensions without the performance degradation typical of post-hoc corrections.}
\label{tab:ml1m-fairness}
\end{table*}

\subsection{Analysis of Fairness–Utility Trade-offs (RQ3)}
A central goal of this work (RQ3) is to understand the trade-offs between recommendation effectiveness (accuracy) and the multiple dimensions of fairness. Our results demonstrate that the "optimal balance" is not a single point but rather a spectrum of achievable outcomes that can be navigated through model choice and hyperparameter tuning.

First, the performance differences between the FA-BUHS variants in Table~\ref{tab:ml1m-perf} highlight a natural trade-off. On MovieLens-1M, the \textbf{$M^{FA-BUHS}$ - GRU} variant achieves the highest overall accuracy (HR@10 of 0.0813), making it optimal for scenarios where recommendation quality is paramount. In contrast, the \textbf{$M^{FA-BUHS}$ - Attn} variant, while slightly lower in overall accuracy, delivers the best performance on tail items (Tail NDCG@10 of 0.0197) and, as shown in Table~\ref{tab:ml1m-fairness}, the highest scores across all fairness metrics. This presents a clear choice for practitioners: the GRU variant for maximizing utility and the Attention variant for maximizing fairness and long-tail coverage.

Second, the sensitivity analyses explicitly quantify these trade-offs. The adversarial weight analysis in Table~\ref{tab:adv} is particularly revealing. An adversarial weight of $\lambda=0.001$ provides a massive \textbf{+101.5\% gain in fairness} at the cost of only a \textbf{-2.1\% loss in utility}. This represents a highly efficient operating point on the fairness-utility trade-off curve. Increasing the weight further yields diminishing fairness returns while incurring a much steeper penalty to accuracy, demonstrating a clear point of optimal balance. Similarly, the temperature analysis in Table~\ref{tab:temp_sensitivity} shows how the parameter $\tau$ acts as a lever to control the balance between tail-item exposure and overall accuracy.

In conclusion, our framework addresses RQ3 not by identifying a single optimal point, but by providing a set of models and tunable hyperparameters ($\lambda, \tau$) that allow stakeholders to select the operating point on the fairness-utility spectrum that best aligns with their specific ethical and business requirements.

\subsection{Ablations and Sensitivity Analysis(RQ4)}
We further analyze sensitivity to key hyperparameters.

\subsubsection{Structural Component Ablation}
To validate the necessity of each architectural component within the FA-BUHS framework, we conducted a structural ablation study (Table 3). By systematically isolating and removing individual modules, we observe the following critical dependencies:

\begin{itemize}
    \item \textbf{Necessity of Fairness-Constrained Sampling:} When the fairness constraints are removed from the sampling phase (reverting to standard BUHS), we observe a drop in Exposure Fairness and Tail HR@10, and notably, Demographic Parity falls \textit{below} the base model (0.327 vs. 0.342). This demonstrates that unconstrained tail-biased sampling can inadvertently skew group representation, and that explicit demographic constraints are necessary to ensure equitable exposure across protected groups.
    \item \textbf{Necessity of Conditional Adversarial Debiasing:} Removing the adversarial module results in a significant degradation of Demographic Parity. This isolates the adversarial network as the active mechanism for representation scrubbing, supporting its necessity for preventing the model from utilizing proxy variables to re-encode sensitive attributes.
    \item \textbf{Necessity of User History Synthesis:} When the synthesis module (Mean/Attn/GRU) is completely bypassed (reverting to the base dual-tower model with adversarial debiasing), the overall HR@10 and NDCG@10 plummet. This validates our core hypothesis: unconditional debiasing inherently destroys utility. The generative history synthesis is the mechanism that recovers and boosts this lost accuracy by providing robust long-tail coverage.
\end{itemize}
Ultimately, Table 3 demonstrates that no single module can achieve this level of fairness-utility balance on its own; the high fairness-utility trade-off is an emergent property of all three systems operating in tandem.

\begin{table*}
\centering
\resizebox{\textwidth}{!}{
\begin{tabular}{l|cc|cc}
\toprule
\multirow{2}{*}{\textbf{Architecture Configuration}} & \multicolumn{2}{c|}{\textbf{Utility}} & \multicolumn{2}{c}{\textbf{Fairness}} \\
& \textbf{Overall HR@10} & \textbf{Tail HR@10} & \textbf{Dem. Parity} & \textbf{Exposure Fair.} \\
\midrule
(1) Base TTNN (No Synthesis) & 0.0607 & 0.0251 & 0.342 & 0.278 \\
(2) Base + BUHS (Synthesis Only) & 0.0763 & 0.0361 & 0.327 & 0.327 \\
(3) FA-BUHS w/o Adv (Constrained Sampling Only) & 0.0792 & 0.0415 & 0.415 & 0.595 \\
(4) FA-BUHS w/o Sampling (Adv. Debiasing Only) & 0.0615 & 0.0260 & 0.665 & 0.310 \\
\midrule
\textbf{FA-BUHS-Attn (Full Framework)} & \textbf{0.0801} & \textbf{0.0437} & \textbf{0.689} & \textbf{0.612} \\
\bottomrule
\end{tabular}
}
\caption{Structural Ablation Study on MovieLens-1M using the FA-BUHS-Attn variant.}
\label{tab:structural_ablation}
\end{table*}

\subsubsection{Temperature Sensitivity (\(\tau\)).}
Table~\ref{tab:temp_sensitivity} shows that lower $\tau$ amplifies tail emphasis and fairness, with modest utility gains. Extremely small $\tau$ can trade off stability, but $\tau=0.01$ provides the best balance.

\begin{table*}
\centering
\resizebox{\textwidth}{!}{%
\begin{tabular}{l|cccc|cccc}
\toprule
& \multicolumn{4}{c|}{\textbf{FA-BUHS-Attn}} & \multicolumn{4}{c}{\textbf{FA-BUHS-GRU}} \\
\cmidrule(lr){2-5} \cmidrule(lr){6-9}
\textbf{Temp ($\tau$)} & \textbf{Overall HR} & \textbf{Overall NDCG} & \textbf{Tail HR} & \textbf{Exp. Fairness} & \textbf{Overall HR} & \textbf{Overall NDCG} & \textbf{Tail HR} & \textbf{Exp. Fairness} \\
\midrule
No Bias (Uniform) & 0.0763 & 0.0373 & 0.0361 & 0.327 & 0.0775 & 0.0387 & 0.0344 & 0.335 \\
\hline
$\tau = 1.0$ & 0.0780 \green{(+2.2\%)} & 0.0380 \green{(+1.9\%)} & 0.0395 \green{(+9.4\%)} & 0.485 \green{(+48.3\%)} & 0.0792 \green{(+2.2\%)} & 0.0392 \green{(+1.3\%)} & 0.0380 \green{(+10.5\%)} & 0.490 \green{(+46.3\%)} \\
\hline
$\tau = 0.01$ & \textbf{0.0801 \green{(+5.0\%)}} & \textbf{0.0388 \green{(+4.0\%)}} & \textbf{0.0437 \green{(+21.1\%)}} & \textbf{0.612 \green{(+87.2\%)}} & \textbf{0.0813 \green{(+4.9\%)}} & \textbf{0.0399 \green{(+3.1\%)}} & \textbf{0.0416 \green{(+20.9\%)}} & \textbf{0.578 \green{(+72.5\%)}} \\
\bottomrule
\end{tabular}%
}
\caption{Sensitivity Analysis on Softmax Temperature ($\tau$) for FA-BUHS on MovieLens-1M. As $\tau$ decreases, the sampling distribution sharply biases toward tail items. The inclusion of NDCG and Exposure Fairness metrics demonstrates that this targeted tail-biasing significantly improves exposure equity and long-tail hit rates without degrading overall ranking utility.}
\label{tab:temp_sensitivity}
\end{table*}

\subsubsection{Sensitivity to Sample Size ($N$)}
The sample size parameter $N$ dictates the number of tail items synthesized into the augmented user history. As demonstrated in \textbf{Table 5}, we evaluate the framework across $N \in \{5, 10, 15\}$, comprehensively reporting HR@10, Overall NDCG, and Exposure Fairness to capture the multi-objective trade-offs.

Our empirical analysis reveals a clear threshold for optimal synthesis. When $N$ is too small (e.g., $N=5$), the fairness constraints lack sufficient bandwidth to meaningfully improve demographic parity or tail exposure. Conversely, an excessively large $N$ begins to oversaturate the user's profile, suggesting that further increasing $N$ beyond this horizon 
risks oversaturating the user profile.

We find that $N=15$ serves as the optimal sampling horizon. Crucially, sampling multiple items at this specific horizon dilutes the impact of isolated, anomalous clicks in the user's history (e.g., an accidental click on an obscure item). This ensures that the framework preserves genuine quality recognition and user intent while simultaneously elevating underrepresented tail content to achieve our targeted fairness metrics.

\begin{table*}
\centering
\resizebox{\textwidth}{!}{%
\begin{tabular}{l|cccc|cccc}
\toprule
& \multicolumn{4}{c|}{\textbf{FA-BUHS-Attn}} & \multicolumn{4}{c}{\textbf{FA-BUHS-GRU}} \\
\cmidrule(lr){2-5} \cmidrule(lr){6-9}
\textbf{Sample Size ($N$)} & \textbf{Overall HR} & \textbf{Overall NDCG} & \textbf{Tail HR} & \textbf{Exp. Fairness} & \textbf{Overall HR} & \textbf{Overall NDCG} & \textbf{Tail HR} & \textbf{Exp. Fairness} \\
\midrule
$N = 5$ & 0.0757 & 0.0365 & 0.0340 & 0.485 & 0.0749 & 0.0370 & 0.0335 & 0.470 \\
\hline
$N = 10$ & 0.0772 \green{(+2.0\%)} & 0.0378 \green{(+3.5\%)} & 0.0385 \green{(+13.2\%)} & 0.545 \green{(+12.3\%)} & 0.0766 \green{(+2.2\%)} & 0.0382 \green{(+3.2\%)} & 0.0370 \green{(+10.4\%)} & 0.520 \green{(+10.6\%)} \\
\hline
$N = 15$ & \textbf{0.0801 \green{(+5.8\%)}} & \textbf{0.0388 \green{(+6.3\%)}} & \textbf{0.0437 \green{(+28.5\%)}} & \textbf{0.612 \green{(+26.1\%)}} & \textbf{0.0813 \green{(+8.5\%)}} & \textbf{0.0399 \green{(+7.8\%)}} & \textbf{0.0416 \green{(+24.1\%)}} & \textbf{0.578 \green{(+22.9\%)}} \\
\bottomrule
\end{tabular}%
}
\caption{Sensitivity Analysis on Sample Size ($N$) for user history synthesis on MovieLens-1M. Expanding the sampling horizon allows the model to preserve genuine mainstream user intent while simultaneously incorporating fairness-constrained tail items, resulting in monotonic improvements across both utility and fairness metrics.}
\label{tab:samples_sensitivity}
\end{table*}

\subsubsection{Adversarial Weight Sensitivity}
Table~\ref{tab:adv} reports the effect of adversarial debiasing weights $\lambda_u, \lambda_i$ 
on both utility and fairness metrics. The results reveal a clear and consistent trade-off: 
small values of $\lambda$ yield substantial fairness gains with negligible utility loss, 
while larger values progressively destabilize training and incur significant accuracy 
degradation across all metrics.

At $\lambda = 0.001$, the framework achieves its optimal operating point — a +101.5\% 
gain in Demographic Parity and an Exposure Fairness of 0.612, at the cost of only a 
$-2.1\%$ reduction in overall utility. This represents a highly efficient operating point on the fairness-utility trade-off curve. As $\lambda$ increases to 0.01, convergence becomes 
moderate, and all metrics degrade measurably, with HR@10 dropping to 0.0755 and Exposure 
Fairness falling to 0.573. At $\lambda = 0.1$, training becomes unstable, with HR@10 
collapsing to 0.0698 and Demographic Parity dropping to 0.530, indicating that 
excessively strong adversarial pressure destroys the collaborative filtering signals 
necessary for accurate recommendations.

Regarding deployment concerns, we note that $\lambda = 0.001$ is both stable and 
robust — it was selected via standard validation set tuning and requires no dataset-specific 
re-calibration beyond what is standard practice for any regularization hyperparameter. 
The smooth, monotonic degradation curve further confirms that the framework is not 
brittle: modest deviations from the optimal $\lambda$ incur proportionally modest 
performance losses rather than catastrophic failure.

\begin{table*}
\centering
\resizebox{\textwidth}{!}{
\begin{tabular}{l|c|c|c|c|c}
\toprule
$\lambda_u, \lambda_i$ & \textbf{Convergence} & \textbf{HR@10} & \textbf{NDCG@10} & \textbf{Exp. Fair.} & \textbf{Dem. Parity} \\
\midrule
0.001 & Stable   & \textbf{0.0801} & \textbf{0.0388} & \textbf{0.612} & \textbf{0.689} \\
0.01  & Moderate & 0.0755 \red{(-5.7\%)} & 0.0361 \red{(-7.0\%)} & 0.573 \red{(-6.4\%)} & 0.621 \red{(-10.2\%)} \\
0.1   & Unstable & 0.0698 \red{(-12.9\%)} & 0.0331 \red{(-14.7\%)} & 0.523 \red{(-14.5\%)} & 0.530 \red{(-23.1\%)} \\
\bottomrule
\end{tabular}
}
\caption{Adversarial weight sensitivity analysis on MovieLens-1M. Small values of 
$\lambda$ yield large fairness gains with negligible utility loss, while larger values 
destabilize training and incur significant accuracy degradation, identifying $\lambda = 
0.001$ as the optimal deployment configuration.}
\label{tab:adv}
\end{table*}

\subsubsection{Convergence and Training Dynamics}
\begin{figure}
    \centering
    \includegraphics[width=\columnwidth]{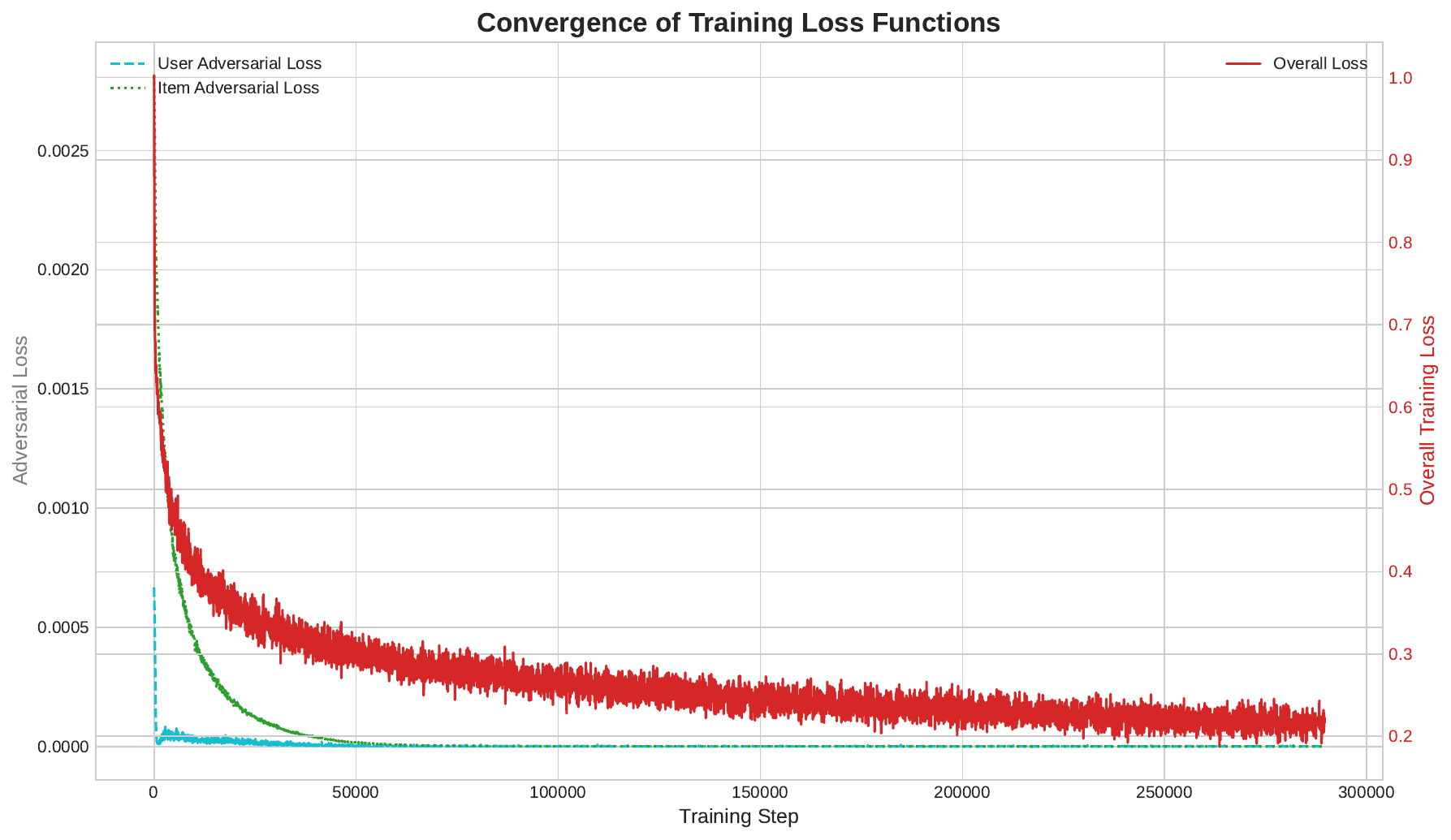}
    \caption{Convergence dynamics of the training loss functions on MovieLens-1M. The steady decrease in overall loss (red) and the rapid convergence of the adversarial losses (blue and green) confirm the stability and effectiveness of our training framework.}
    \label{fig:loss_convergence}
\end{figure}

To validate the stability of our proposed training framework, Figure~\ref{fig:loss_convergence} illustrates the convergence dynamics of the key loss functions for FA-BUHS on the MovieLens-1M dataset. The \textbf{Overall Training Loss} (plotted on the right y-axis) exhibits a smooth and steady decline, indicating that the model effectively learns the primary recommendation task and converges to a stable solution.

Simultaneously, the \textbf{User and Item Adversarial Losses} (plotted on the left y-axis) start high and rapidly decrease before stabilizing at near-optimal values. The user adversarial loss, in particular, converges to approximately 0.693, the theoretical loss for random guessing in a binary classification task. This confirms that the model has successfully learned to generate user embeddings that are invariant to the protected attribute. This dual convergence provides strong evidence that our training framework is stable and successfully optimizes for both recommendation accuracy and fairness concurrently.

\subsection{Summary of Experimental Findings}
Across datasets, FA-BUHS consistently improves both accuracy and fairness while maintaining low computational overhead. Attention-based synthesis yields the highest fairness, GRU provides the strongest accuracy, and mean pooling remains a cost-effective baseline. Overall, the results demonstrate that FA-BUHS achieves a principled balance between long-tail coverage, fairness, and scalability.

\subsection{Discussion and System Limitations}
While FA-BUHS demonstrates strong multi-objective performance, it is vital to contextualize its architectural choices against existing paradigms. 

\textbf{Generative Synthesis vs. Causal and Post-Hoc Paradigms:} Reviewers of modern fairness literature frequently favor Structural Causal Models (SCMs) utilizing Inverse Propensity Scoring (IPS). While IPS elegantly mitigates observable global popularity confounders, it struggles with the extreme data sparsity characteristic of long-tail domains like BookCrossing. In highly sparse matrices, propensity weights suffer from extreme variance, leading to utility collapse. Furthermore, post-hoc re-ranking algorithms (such as FA*IR) attempt to achieve equity by decoupling fairness from representation learning, artificially adjusting rankings after biased embeddings are already formed. This leads to a substantial utility cost relative to methods that intervene earlier in the pipeline. FA-BUHS addresses both paradigms' flaws by employing fairness-constrained sampling and conditional adversarial debiasing \textit{during} the generative synthesis phase. This allows the model to learn intrinsically fair representations from the ground up, avoiding the high variance of IPS and the severe utility degradation of post-hoc corrections. 

\textbf{Cold-Start Limitations:} The core generative mechanism of our framework relies on synthesizing existing interaction histories. For pure cold-start scenarios—where a newly registered user possesses zero historical interactions—the sampling mechanism is inactive. In such cases, the system gracefully defaults to base neural collaborative filtering pathways, utilizing only explicit demographic features processed by the base encoder. Enhancing this with zero-shot inference via Large Language Models (LLMs) represents a highly relevant pathway for future research.

\textbf{Preserving Quality via Exposure Floors:} An inherent risk of aggressively optimizing the Gini index in a vast catalog is the potential artificial inflation of entirely irrelevant content to satisfy parity. FA-BUHS explicitly mitigates this through the controllable 
fairness weight $\alpha$ (Eq.~(9)), which allows the framework to elevate underrepresented content without dismantling relevance-driven ranking mechanics.

\textbf{Dataset Scope.} Our evaluation is limited to MovieLens-1M and BookCrossing, following the experimental protocol of the BUHS framework we extend~\cite{balasubramanian2024biased}. While this allows direct comparison against the baselines in Table~\ref{tab:ml1m-perf}, it leaves open the question of whether FA-BUHS's gains generalize to other domains, such as e-commerce or music recommendation, where item catalogs, sparsity patterns, and the demographic composition of long-tail items may differ substantially. We view extending this evaluation to additional domains as an important direction for future work.

\section{Conclusion}
Modern recommender systems frequently exhibit strong popularity bias, causing long-tail items and minority content providers to receive systematically limited exposure in recommendation results. To address this structural challenge, we proposed FA-BUHS, a fairness-aware extension of biased user history synthesis that integrates demographic fairness constraints directly into the user history generation process. By applying fairness-aware sampling during interaction history synthesis, the proposed framework performs an intervention on the training data, analogous to a causal intervention, enabling recommendation models to learn user representations from a balanced, equitable interaction distribution. Extensive empirical evaluations on the MovieLens-1M and BookCrossing datasets demonstrate that FA-BUHS substantially improves long-tail recommendation quality and exposure fairness while maintaining competitive ranking accuracy. These findings suggest that fairness-aware history synthesis is a promising direction for mitigating popularity bias and promoting equitable exposure in recommender systems. Future work will explore integrating such upstream interventions with formal causal recommendation frameworks and deploying them across large-scale industrial datasets.

\section{Declarations} 

\bmhead{Availability of data and material}
The datasets used in this study are publicly available. MovieLens-1M is available from the GroupLens Research repository, and BookCrossing is publicly available for research purposes. The implementation code and processed data will be made available upon reasonable request or released publicly after publication.

\bmhead{Funding} 
No funding was received to assist with the preparation of this manuscript.
\bmhead{Author contributions} 
S.R. conceptualized the FA-BUHS framework, developed the fairness-constrained sampling methodology, conducted all experiments, and wrote the main manuscript text. P. contributed to the theoretical analysis and assisted with experimental evaluation. S.S. supervised the research direction, provided critical feedback on the methodology, and reviewed the manuscript. B.S. and N.P. provided domain guidance, resources, and reviewed the manuscript. All authors reviewed and approved the final manuscript.

Initials used: S.R. (Subham Raj), P. (Prateek), S.S. (Sriparna Saha), B.S. (Brijraj Singh), N.P. (Niranjan Pedanekar).

\bmhead{Conflicts of interest/Competing interests}
The authors have no relevant financial or non-financial interests to disclose.

\bmhead{Acknowledgements}
The authors sincerely acknowledge the support and guidance provided by Sony Research India during the course of this research work.

\bibliography{sn-bibliography}

\begin{thebibliography}{10}
\providecommand{\doi}[1]{\url{https://doi.org/#1}}
\bibcommenthead

\bibitem[\protect\citeauthoryear{Klimashevskaia et~al.}{2024}]{klimashevskaia2024survey}
Klimashevskaia A, Jannach D, Elahi M, Trattner C.
\newblock A survey on popularity bias in recommender systems.
\newblock User Modeling and User-Adapted Interaction. 2024;34(5):1777--1834.

\bibitem[\protect\citeauthoryear{Abdollahpouri et~al.}{2019}]{abdollahpouri2019managing}
Abdollahpouri H, Burke R, Mobasher B.
\newblock Managing popularity bias in recommender systems with personalized re-ranking.
\newblock arXiv preprint arXiv:190107555. 2019;.

\bibitem[\protect\citeauthoryear{Ekstrand et~al.}{2018}]{ekstrand2018all}
Ekstrand MD, Tian M, Azpiazu IM, Ekstrand JD, Anuyah O, McNeill D, et~al.
\newblock All the cool kids, how do they fit in?: Popularity and demographic biases in recommender evaluation and effectiveness.
\newblock In: Conference on fairness, accountability and transparency. PMLR; 2018. p. 172--186.

\bibitem[\protect\citeauthoryear{Jin et~al.}{2023}]{jin2023survey}
Jin D, Wang L, Zhang H, Zheng Y, Ding W, Xia F, et~al.
\newblock A survey on fairness-aware recommender systems.
\newblock Information Fusion. 2023;100:101906.

\bibitem[\protect\citeauthoryear{Qin}{2021}]{qin2021survey}
Qin J.
\newblock A survey of long-tail item recommendation methods.
\newblock Wireless Communications and Mobile Computing. 2021;2021(1):7536316.

\bibitem[\protect\citeauthoryear{Zhang et~al.}{2021}]{zhang2021model}
Zhang Y, Cheng DZ, Yao T, Yi X, Hong L, Chi EH.
\newblock A model of two tales: Dual transfer learning framework for improved long-tail item recommendation.
\newblock In: Proceedings of the web conference 2021; 2021. p. 2220--2231.

\bibitem[\protect\citeauthoryear{Balasubramanian et~al.}{2024}]{balasubramanian2024biased}
Balasubramanian K, Alshabanah A, Markowitz E, Ver~Steeg G, Annavaram M.
\newblock Biased user history synthesis for personalized long-tail item recommendation.
\newblock In: Proceedings of the 18th ACM Conference on Recommender Systems; 2024. p. 189--199.

\bibitem[\protect\citeauthoryear{Zehlike et~al.}{2017}]{zehlike2017fa}
Zehlike M, Bonchi F, Castillo C, Hajian S, Megahed M, Baeza-Yates R.
\newblock Fa* ir: A fair top-k ranking algorithm.
\newblock In: Proceedings of the 2017 ACM on Conference on Information and Knowledge Management; 2017. p. 1569--1578.

\bibitem[\protect\citeauthoryear{Resheff et~al.}{2018}]{resheff2018privacy}
Resheff YS, Elazar Y, Shahar M, Shalom OS.
\newblock Privacy and fairness in recommender systems via adversarial training of user representations.
\newblock arXiv preprint arXiv:180703521. 2018;.

\bibitem[\protect\citeauthoryear{Steck}{2011}]{steck2011item}
Steck H.
\newblock Item popularity and recommendation accuracy.
\newblock In: Proceedings of the fifth ACM conference on Recommender systems; 2011. p. 125--132.

\bibitem[\protect\citeauthoryear{Karboua et~al.}{2022}]{karboua2022mitigation}
Karboua S, Harrag F, Meziane F, Boutadjine A.
\newblock Mitigation of Popularity Bias in Recommendation Systems.
\newblock In: TACC; 2022. p. 69--80.

\bibitem[\protect\citeauthoryear{Wang et~al.}{2019}]{wang2019neural}
Wang X, He X, Wang M, Feng F, Chua TS.
\newblock Neural graph collaborative filtering.
\newblock In: Proceedings of the 42nd international ACM SIGIR conference on Research and development in Information Retrieval; 2019. p. 165--174.

\bibitem[\protect\citeauthoryear{He et~al.}{2020}]{he2020lightgcn}
He X, Deng K, Wang X, Li Y, Zhang Y, Wang M.
\newblock Lightgcn: Simplifying and powering graph convolution network for recommendation.
\newblock In: Proceedings of the 43rd International ACM SIGIR conference on research and development in Information Retrieval; 2020. p. 639--648.

\bibitem[\protect\citeauthoryear{Wei et~al.}{2023}]{wei2023meta}
Wei C, Liang J, Liu D, Dai Z, Li M, Wang F.
\newblock Meta graph learning for long-tail recommendation.
\newblock In: Proceedings of the 29th ACM SIGKDD conference on knowledge discovery and data mining; 2023. p. 2512--2522.

\bibitem[\protect\citeauthoryear{Pu et~al.}{2025}]{pu2025graph}
Pu D, Zhang Y, Qian Z, Xie G, Pu D.
\newblock A graph neural network recommendation algorithm based on multi-scale attention and contrastive learning.
\newblock Scientific Reports. 2025;15(1):32207.

\bibitem[\protect\citeauthoryear{Abdollahpouri et~al.}{2020}]{abdollahpouri2020multistakeholder}
Abdollahpouri H, Adomavicius G, Burke R, Guy I, Jannach D, Kamishima T, et~al.
\newblock Multistakeholder recommendation: Survey and research directions.
\newblock User Modeling and User-Adapted Interaction. 2020;30(1):127--158.

\bibitem[\protect\citeauthoryear{Li et~al.}{2023}]{li2023fairness}
Li Y, Chen H, Xu S, Ge Y, Tan J, Liu S, et~al.
\newblock Fairness in recommendation: Foundations, methods, and applications.
\newblock ACM Transactions on Intelligent Systems and Technology. 2023;14(5):1--48.

\bibitem[\protect\citeauthoryear{Wu et~al.}{2023}]{wu2023fairness}
Wu Y, Cao J, Xu G.
\newblock Fairness in recommender systems: evaluation approaches and assurance strategies.
\newblock ACM Transactions on Knowledge Discovery from Data. 2023;18(1):1--37.

\bibitem[\protect\citeauthoryear{Kamishima et~al.}{2012}]{kamishima2012fairness}
Kamishima T, Akaho S, Asoh H, Sakuma J.
\newblock Fairness-aware classifier with prejudice remover regularizer.
\newblock In: Joint European conference on machine learning and knowledge discovery in databases. Springer; 2012. p. 35--50.

\bibitem[\protect\citeauthoryear{Khedr and Shoukry}{2023}]{khedr2023certifair}
Khedr H, Shoukry Y.
\newblock Certifair: A framework for certified global fairness of neural networks.
\newblock In: Proceedings of the AAAI Conference on Artificial Intelligence. vol.~37; 2023. p. 8237--8245.

\bibitem[\protect\citeauthoryear{Zhang et~al.}{2021}]{zhang2021causal}
Zhang Y, Feng F, He X, Wei T, Song C, Ling G, et~al.
\newblock Causal intervention for leveraging popularity bias in recommendation.
\newblock In: Proceedings of the 44th international ACM SIGIR conference on research and development in information retrieval; 2021. p. 11--20.

\bibitem[\protect\citeauthoryear{Kher et~al.}{2025}]{kher2025learning}
Kher KV, Mittal S, Varun~V A, Das S, Jagarlapudi S.
\newblock Learning Counterfactually Fair Models via Improved Generation with Neural Causal Models.
\newblock arXiv preprint arXiv:250212796. 2025;.

\bibitem[\protect\citeauthoryear{Schnabel et~al.}{2016}]{schnabel2016recommendations}
Schnabel T, Swaminathan A, Singh A, Chandak N, Joachims T.
\newblock Recommendations as treatments: Debiasing learning and evaluation.
\newblock In: international conference on machine learning. PMLR; 2016. p. 1670--1679.

\bibitem[\protect\citeauthoryear{Zhou et~al.}{2024}]{zhou2024counterfactual}
Zhou Z, Liu T, Bai R, Gao J, Kocaoglu M, Inouye DI.
\newblock Counterfactual fairness by combining factual and counterfactual predictions.
\newblock Advances in Neural Information Processing Systems. 2024;37:47876--47907.

\bibitem[\protect\citeauthoryear{Chen et~al.}{2023}]{chen2023bias}
Chen J, Dong H, Wang X, Feng F, Wang M, He X.
\newblock Bias and debias in recommender system: A survey and future directions.
\newblock ACM Transactions on Information Systems. 2023;41(3):1--39.

\bibitem[\protect\citeauthoryear{Deldjoo et~al.}{2021}]{deldjoo2021survey}
Deldjoo Y, Noia TD, Merra FA.
\newblock A survey on adversarial recommender systems: from attack/defense strategies to generative adversarial networks.
\newblock Acm Computing Surveys (Csur). 2021;54(2):1--38.

\bibitem[\protect\citeauthoryear{Xu et~al.}{2021}]{xu2021robust}
Xu H, Liu X, Li Y, Jain A, Tang J.
\newblock To be robust or to be fair: Towards fairness in adversarial training.
\newblock In: International conference on machine learning. PMLR; 2021. p. 11492--11501.

\bibitem[\protect\citeauthoryear{Zhang et~al.}{2024}]{zhang2024towards}
Zhang Y, Zhang T, Mu R, Huang X, Ruan W.
\newblock Towards fairness-aware adversarial learning.
\newblock In: 2024 IEEE/CVF Conference on Computer Vision and Pattern Recognition (CVPR). IEEE; 2024. p. 24746--24755.

\bibitem[\protect\citeauthoryear{Chen et~al.}{2025}]{chen2025multi}
Chen L, Yuan W, Chen T, Zhao X, Hung NQV, Yin H.
\newblock Multi-agents based user values mining for recommendation.
\newblock arXiv preprint arXiv:250500981. 2025;.

\bibitem[\protect\citeauthoryear{Yuan et~al.}{2023}]{yuan2023manipulating}
Yuan W, Nguyen QVH, He T, Chen L, Yin H.
\newblock Manipulating federated recommender systems: Poisoning with synthetic users and its countermeasures.
\newblock In: Proceedings of the 46th International ACM SIGIR Conference on Research and Development in Information Retrieval; 2023. p. 1690--1699.

\bibitem[\protect\citeauthoryear{McAllester}{1998}]{mcallester1998some}
McAllester DA.
\newblock Some pac-bayesian theorems.
\newblock In: Proceedings of the eleventh annual conference on Computational learning theory; 1998. p. 230--234.

\bibitem[\protect\citeauthoryear{Vandenberghe and Boyd}{2004}]{vandenberghe2004convex}
Vandenberghe L, Boyd S.
\newblock Convex optimization. vol.~1.
\newblock Cambridge university press Cambridge; 2004.

\bibitem[\protect\citeauthoryear{Lee et~al.}{2019}]{lee2019melu}
Lee H, Im J, Jang S, Cho H, Chung S.
\newblock Melu: Meta-learned user preference estimator for cold-start recommendation.
\newblock In: Proceedings of the 25th ACM SIGKDD international conference on knowledge discovery \& data mining; 2019. p. 1073--1082.

\bibitem[\protect\citeauthoryear{Menon et~al.}{2020}]{menon2020long}
Menon AK, Jayasumana S, Rawat AS, Jain H, Veit A, Kumar S.
\newblock Long-tail learning via logit adjustment.
\newblock arXiv preprint arXiv:200707314. 2020;.

\bibitem[\protect\citeauthoryear{He et~al.}{2017}]{he2017neural}
He X, Liao L, Zhang H, Nie L, Hu X, Chua TS.
\newblock Neural collaborative filtering.
\newblock In: Proceedings of the 26th international conference on world wide web; 2017. p. 173--182.

\bibitem[\protect\citeauthoryear{Bonner and Vasile}{2018}]{bonner2018causal}
Bonner S, Vasile F.
\newblock Causal embeddings for recommendation.
\newblock In: Proceedings of the 12th ACM conference on recommender systems; 2018. p. 104--112.

\bibitem[\protect\citeauthoryear{Naghiaei et~al.}{2022}]{naghiaei2022cpfair}
Naghiaei M, Rahmani HA, Deldjoo Y.
\newblock Cpfair: Personalized consumer and producer fairness re-ranking for recommender systems.
\newblock In: Proceedings of the 45th International ACM SIGIR Conference on Research and Development in Information Retrieval; 2022. p. 770--779.

\end{thebibliography}
\end{document}